\documentclass{emulateapj}
  \usepackage{epsfig}

\shortauthors{Stanek \ea}
\shorttitle{Halos in Millennium Gas Simulations: Multivariate Scaling Relations}

\newcommand{\ea}{et al.}
\newcommand{\lta}{\lesssim}

\newcommand{\kms}{\>{\rm km}\,{\rm s}^{-1}}

\newcommand{\mpc}{\>{\rm Mpc}}

\newcommand{\kev}{\>{\rm keV}}
\newcommand{\cgsflux}{\rm\ erg\ s^{-1}\ cm^{-2}}

\newcommand{\msun}{\>{\rm M_{\odot}}}
\newcommand{\msol}{\>{\rm M_{\odot}}}
\newcommand{\hinv}{\>{h^{-1}}}

\newcommand{\bdm}{\begin{displaymath}} 
\newcommand{\edm}{\end{displaymath}}
\newcommand{\beq}{\begin{equation}} 
\newcommand{\eeq}{\end{equation}} 
\newcommand{\bit}{\begin{itemize}} 
\newcommand{\eit}{\end{itemize}} 
\newcommand{\ben}{\begin{enumerate}} 
\newcommand{\een}{\end{enumerate}}

\newcommand{\GO}{\texttt{GO} }
\newcommand{\PH}{\texttt{PH} }
\newcommand{\ficm}{\hbox{$f_{\rm ICM}$}}
\newcommand{\elbol}{\hbox{$L_{\rm bol}$}}

\newcommand{\tsl}{\hbox{$T_{\rm sl}$}}
\newcommand{\sigmadm}{\hbox{$\sigma_{\rm DM}$}}
\newcommand{\sigmagas}{\hbox{$\sigma_{\rm gas}$}}

\newcommand{\betagas}{\hbox{$\beta_{\rm gas}$}}

\newcommand{\maxbcg}{\hbox{\tt maxbcg}}

\begin{document}

\title{Massive Halos in Millennium Gas Simulations: Multivariate Scaling Relations}

\author{R. Stanek}
\affil{Department of Astronomy, University of Michigan, 500 Church St., Ann Arbor, MI  48109}
\email{rstanek@umich.edu}

\author{E. Rasia}
\affil{{\sl Chandra} Fellow, Department of Astronomy and Michigan Society of Fellows, University of Michigan, 500 Church St., Ann Arbor, MI  48109}

\author{A. E. Evrard} 
\affil{Departments of Physics and Astronomy and Michigan Center for Theoretical Physics, University of Michigan,  Ann Arbor, MI  48109}

\author{F. Pearce}
\affil{School of Physics and Astronomy, University of Nottingham, Nottingham, NG7 2RD, UK }

\author{L. Gazzola}
\affil{School of Physics and Astronomy, University of Nottingham, Nottingham, NG7 2RD, UK }

%%%%%%%%%%%%%%%
% Start the abstract on a fresh page
%%%%%%%%%%%%%%%

\clearpage

%%%%%%%%%%%%%%%
% Abstract
%%%%%%%%%%%%%%%

\begin{abstract} 

The joint likelihood of observable cluster signals reflects the astrophysical evolution of the coupled baryonic and dark matter components in massive halos, and its knowledge will enhance cosmological parameter constraints in the coming era of large, multi-wavelength cluster surveys.  We present a  computational study of intrinsic covariance in cluster properties using halo populations derived from Millennium Gas Simulations (MGS).  The MGS are re-simulations of the original $500 \hinv \mpc$ Millennium Simulation performed with gas dynamics under two different physical treatments: shock heating driven by gravity only ($\GO$) and a second treatment with cooling and preheating ($\PH$).   We examine relationships among structural properties and observable X-ray and Sunyaev-Zel'dovich (SZ) signals for samples of thousands of halos with $M_{200} \ge 5 \times 10^{13} \hinv\msun$ and $z < 2$.     While the X-ray scaling behavior of \PH model halos at low-redshift offers a good match to local clusters, the model exhibits non-standard features testable with larger surveys, including weakly running slopes in hot gas observable--mass relations and $\sim 10\%$ departures from self-similar redshift evolution for $10^{14} \hinv\msun$ halos at redshift $z \sim 1$.    We find that the form of the joint likelihood of signal pairs is generally well-described by a multivariate, log-normal distribution, especially in the \PH case which exhibits less halo substructure than the \GO model.  At fixed mass and epoch, joint deviations of signal pairs display mainly positive correlations, especially the thermal SZ effect paired with either hot gas fraction ($r=0.88/0.69$ for $\PH/\GO$ at $z=0$) or X-ray temperature ($r=0.62/0.83$).    The levels of variance in X-ray luminosity, temperature and gas mass fraction are sensitive to the physical treatment, but offsetting shifts in the latter two measures maintain a fixed $12\%$ scatter in the integrated SZ signal under both gas treatments.  We discuss halo mass selection by signal pairs, and find a minimum mass scatter of $4\%$ in the \PH model by combining thermal SZ and gas fraction measurements.   
\end{abstract}

%%%%%%%%%%%%%%%%%%%%%%%%%%%%%%%%%%%%%%%%%%%%%%%%%%%%%%%%%%%%%%%%%%%%%
%%%%%%%%%%%%%%%%%%%%%%%%%%%%%%%%%%%%%%%%%%%%%%%%%%%%%%%%%%%%%%%%%%%%%
%%%%%%%%%%%%%%%%%%%%%%%%%%%%%%%%%%%%%%%%%%%%%%%%%%%%%%%%%%%%%%%%%%%%%
%%%%%%%%%%%%%%%%%%%%%%%%%%%%%%%%%%%%%%%%%%%%%%%%%%%%%%%%%%%%%%%%%%%%%

\section{Introduction}\label{intro}

Accurate cosmology using surveys of clusters of galaxies requires a robust description of the relations between observed cluster signals and underlying halo mass.  Even without strong prior knowledge of the mass-signal relation, cluster counts, in combination with other probes, add useful constraining power to cosmological parameters \citep{cunha:09a}.   However, significant improvements can be realized when the error in mass variance is known \citep{limaHu:05, cunha:09b}.   Improvements can also be gained  by extending the model to multiple observed signals \citep{cunha:08}, especially when an underlying physical model can effectively reduce the dimensionality of the parameter sub-space associated with the model \citep{younger:06}.   The coming era of multiple observable signals from combined surveys in optical, sub-mm and X-ray wavebands invites a more holistic approach to modeling multi-wavelength signatures of clusters.   

Signal covariance characterizes survey selection, in terms of mass and additional observables.  
For the case of X-ray selected samples, \cite{nord:08} demonstrate that luminosity--temperature covariance can mimic apparent evolution in the luminosity--mass relation under analysis that combines deep, X-ray-flux limited samples with local, shallow ones. 

Employing a selection observable with small mass variance minimizes such errors.  Recent work has shown that the total gas thermal energy, $Y$, observable via an integrated Sunyaev-Zeldovich (SZ) effect \citep{carlstrom:04} or via X-ray imaging and spectroscopy, is a signal that scales as a power-law in mass with only $\sim 15\%$ scatter \citep{white:02, kravtsov:06, maughan:07, ohara:07, zhang:08, jeltema:08}.   However, unbiased estimates of the mass selection function for $Y$ or any other signal requires accurate knowledge of how the signal--mass scaling relation evolves with redshift.   The redshift behavior of signals is generally not well known empirically, although recent work has begun to probe evolution in X-ray signals to $z \sim 1$ \citep{maughan:06,vikhlinin:08}.  Emerging samples from wide-area SZ surveys should dramatically improve this situation.  

One can address signal--mass covariance using hydrostatic, virial or lensing mass estimates from observations, but several sources of systematic and statistical error challenge this approach.   For hydrostatic masses, early gas dynamic simulations \citep{evrard:90,nfw, thomas:97} suggested that turbulent gas motions drove hydrostatic masses to underestimate true values by $\sim 20\%$.  More sophisticated recent models, with a factor thousand improvement in mass resolution,  demonstrate this effect at a similar level in the mean, with $\sim 15\%$ scatter among individual systems \citep{rasia:06, nagai:06, jeltema:08}.  Cluster masses can also be measured by the shear induced on background galaxies due to gravitational lensing.  With this method, individual  cluster masses have mass uncertainties $\sim 20\%$ due to cosmic web confusion \citep{hoekstra:03, deputter:05}, but large samples can reduce the uncertainty in the mean.  

The mean scaling behavior of samples binned in some selection signal offers another empirical path to measuring covariance.   
Non-zero covariance between the selection and an independent, follow-up signal implies that the selection-binned scaling relation of the followup signal with mass need not match that signal's intrinsic mass scaling.  Comparison of scaling relations from differently selected samples thereby offers insight into covariance.   \cite{rykoff:08} offer a first attempt at this exercise for X-ray luminosity and optical richness using the optically-selected SDSS \maxbcg\  sample \citep{koester:07}.  The sample contains $\sim 13,000$ clusters for which weak lensing mass estimates have been made by stacking the shear of richness-binned sub-samples \citep{sheldon:07,johnston:07}.   \cite{rykoff:08} stack Rosat All-Sky Survey data \citep{voges:99} in the same \maxbcg\  richness bins, and find that the mean X-ray luminosity--mass relation derived with richness binning is consistent at the $\sim 2 \sigma$ level with relations derived solely from X--ray data \citep{reiprich:02, stanek:06}.  

A theoretical approach to studying cluster covariance is to realize populations via numerical simulation.  While high resolution treatment of astrophysical processes, including star formation, supernova and AGN feedback, galactic winds and thermal conduction have been included in recent simulations \citep{dolag:05, borgani:06, kravtsov:06, sijacki:08, puchwein:08}, the computational expense has limited sample sizes to typically a few dozen objects.   A detailed study of population covariance requires larger sample sizes, as can be generated by lower resolution simulations of large cosmological volumes using a more limited physics treatment \citep{bryan:98, hallman:06, gottlober:07}.  

We take the latter approach in this paper, focusing on the bulk properties of massive halos identified the Millennium Gas Simulations (MGS), a pair of resimulations of the original $500 \hinv$ Mpc Millennium run \citep{springel:05}, each with $10^9$ total particles, half representing gas and half dark matter.  The pair of runs use different treatments for the gas physics --- a gravity-only ({\texttt{GO}}) simulation, sometimes called ``adiabatic'',  in which entropy is increased via shocks, and a simulation with cooling and preheating  ({\texttt{PH}}).  The former ignores galaxies as both a sink for baryons and a source of feedback for the hot intracluster medium (ICM).  The latter also ignores the mass fraction contribution of galaxies, but it approximates the feedback effects of galaxy formation by a single parameter, an entropy level imposed as a floor at high redshift  \citep{evrard:91, kaiser:91, bialek:01, kay:07, gottlober:07}.  Our study focuses on samples of $\sim 5000$ halos with mass $M > 5 \times 10^{13} \hinv \msun$ examined at multiple epochs covering the redshift range $0 < z < 2$.  

The paper is organized as follows:  in Section \ref{sec:sims} we discuss the 
details of the Millennium Gas Simulations and our halo finding approach. 
  In Section \ref{sec:scaling}, we present mean scaling relation behavior in both mass and redshift.  We then turn to 
 covariance about the mean scaling relations in Section \ref{sec:covar}.  Unless otherwise noted, our units of 
mass are $10^{14} \hinv \msun$ and halo mass is defined within a sphere encompassing a density contrast $\Delta_c = 200$ times the critical density.  
 
%%%%%%%%%%%%%%%%%%%%%%%%%%%%%%%%%%%%%%%%%%%%%%%%%%%%%%%%%%%%%%%%%%%%%
%%%%%%%%%%%%%%%%%%%%%%%%%%%%%%%%%%%%%%%%%%%%%%%%%%%%%%%%%%%%%%%%%%%%%
\section{Simulations} \label{sec:sims}
 
%%%%%%%%%%%%%%%%%%%%%%%%%%%%%%%%%%%%%%%%%%%%%%%%%%%%%%%%%%%%%%%%%%%%%
\subsection{Millennium Gas Simulations} \label{sec:mgs}

The Millennium Gas Simulations (hereafter MGS) are a pair of resimulations of the original 
 Millennium \citep{springel:05}, a high-resolution, dark-matter-only simulation 
of a $500 \hinv$ Mpc volume. The simulations were run with GADGET-2,   
treating the gas dynamics with smoothed particle hydrodynamics (SPH) \citep{gadget2}.  
As described in \cite{hartley:08}, the MGS use the initial conditions of the Millennium simulation, with $5 \times 10^8$ dark matter particles, each of mass $1.42 \times 10^{10} \hinv \msun$, and $5 \times 10^8$ SPH gas particles, each of 
mass $3.12 \times 10^9 \hinv \msun$, resulting in a mass resolution about 20 times coarser than 
the original Millennium N-body simulation.  The gravitational softening length is $25 \hinv$ kpc.  The 
cosmological parameters match the original: $\Omega_m = 1 - \Omega_\Lambda = 0.25$, $\Omega_b = 0.045$, $h = 0.73$, and 
$\sigma_8 = 0.9$.   While some differences between the simulations are expected due to the difference 
in mass resolution and gravitational softening length,  \cite{hartley:08} verify the positions of 
dark matter halos to within $50 \hinv$ kpc between the original Millennium and the MGS.
The value of $\sigma_8$ is higher than the WMAP3 value \citep{spergel:06}, but we 
do not expect it to strongly affect the results of this paper.  
% Shifting $\sigma_8$ to a lower value will merely move the halo populations along the axis of the scaling relations.

In this paper, we consider two models of the MGS:  a \GO simulation where the 
only source of gas entropy change is from shocks, and a \PH simulation with 
preheating and cooling along with shock heating.  The \GO simulation is useful 
as a base model that can be easily compared to previous hydrodynamic simulations 
of galaxy clusters.  Furthermore, comparing gravity-only simulations to observations highlights 
the cluster properties that are strongly affected by astrophysical processes beyond gravitational heating.

While adiabatic simulations match the self-similar prediction, $L \sim T^2$, for the X-ray luminosity--temperature relation, 
observations show a steeper slope \citep{arnaud:99, osmond:04}.  Preheating, the assumption of an elevated initial gas entropy at high redshift, was introduced by \cite{evrard:91} and \cite{kaiser:91} as a means to resolve the discrepancy in shape between the observed X-ray luminosity function and that expected from self-similar scaling of the cosmic mass function.   
The \PH simulation is tuned to match X-ray observations of clusters at redshift zero, particularly the luminosity-temperature (L-T) relation \citep{hartley:08}.  The preheating is achieved by boosting the entropy of 
every gas particle to 200 keV cm$^2$ at redshift $z=4$.  Although the preheating dominates in the \PH 
simulation, there is also cooling based on the cooling function of \cite{sutherland:93}.   Fewer than 
$2\%$ of the baryons are converted to stars, however, and star formation is essentially halted by the 
preheating at $z=4$.  

This simple model certainly does not capture all of the complex astrophysical effects associated with star and supermassive black hole formation in clusters.  The central entropy structure in local clusters is distinctly bimodal, with roughly half the population centered near the 200 keV cm$^2$ value used in the \PH and the remainder centered on an entropy level a factor of ten smaller.   
However, the latter reflects potentially cyclical AGN feeding and feedback \citep{voit:02,sijacki:07,puchwein:08} that strongly influences only a small mass fraction of the total cluster gas.   Observational evidence of ubiquitous galactic winds at high and moderate redshifts \citep{pettini:01,weiner:08} suggest that most of the heating of the ICM occurs at high redshift.  Further support for fast feedback comes from red sequence galaxies extending to redshift $z=1.4$ in the Spitzer/IRAC Shallow Survey \citep{eisenhardt:08}, the colors of which are consistent with passive evolution of a burst of star formation at redshift $z \sim 3-4$.  

%%%%%%%%%%%%%%%%%%%%%%%%%%%%%%%%%%%%%%%%%%%%%%%%%%%%%%%%%%%%%%%%%%%%%
\subsection{Halo Catalog} \label{sec:grp}

We identify halos in the simulation as spherical regions, centered on filtered density peaks, 
that encompass an average density of $\Delta_c \rho_c(z)$, 
where $\rho_c(z)$ is the critical density of the universe.  Both dark matter and baryons are included in the density measurement.  
We use halos identified at an overdensity of $\Delta_c = 200$ for most of 
our analysis.  Halo centers were identified with an $N^{th}$ nearest 
neighbor approach, which approximates the local density by calculating 
the distance to the $32^{nd}$  nearest dark matter particle.  The groupfinder 
begins with the dark matter particle with the highest local density, and works 
outward in radius particle by particle (including gas and dark matter) until 
the interior mean density is $200 \rho_c(z)$.  The algorithm then identifies 
the densest dark matter particle not already in a halo, and continues iteratively 
until all overdense regions with more than 100 particles have been identified.  
Overlapping halos are permitted; however, the center of mass of a halo may not be in another 
halo.  

\begin{figure}
\plotone{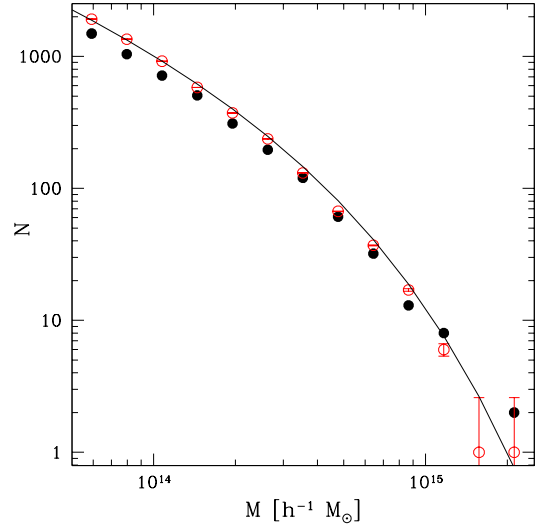}
\caption{Differential counts of halos versus total mass at redshift zero for the \PH simulations (filled, black points), 
the \GO simulation (open, red points), and the prediction from the Tinker mass 
function \citep{tinker:08} (solid, black line). 
\label{fig:massfunc}}
\vspace{0.20 truecm}
\end{figure}

At redshift zero, in the \PH simulation we have approximately 220,000
 halos with at least 100 particles, and 
4474 over a mass cut of $M > 5 \times 10^{13} \hinv M_\odot$.  These numbers are 
higher in the \GO simulation, with approximately 370,000 halos with at least 
100 particles, and 5612 over the mass cut of $M > 5 \times 10^{13} \hinv M_\odot$. 
In Figure \ref{fig:massfunc}, we plot differential halo counts as a function of total mass from the \GO and \PH simulations, as well as the prediction from the Tinker 
mass function (TMF) \citep{tinker:08}.   As discussed in \cite{stanek:09}, preheating causes a decrease in total halo mass of up to $\sim 15\%$ relative to the GO treatment, with the largest effects at lower masses and higher redshifts.  While the \GO halo space density matches the TMF expectations well, the number of halos in the PH case is lower, especially at lower mass.  

Among the halos over the mass cut of $M > 5 \times 10^{13} \hinv M_\odot$, we 
identify overlapping halos.  For a pair of overlapping halos, we denote the less massive 
halo as a ``satellite'' and the more massive halo as a ``primary'' halo.  For the rest of 
the analysis in this paper, we exclude the satellite halos.  At redshift zero we are 
left with a sample of 4404 halos in the \PH simulation and 5498 halos in the \GO 
simulation.

We repeat the halo finding exercise at all redshifts available for each model.  In the case of PH, we employ a total of 63 outputs extending to a redshift of two.  For the GO, we analyze only a subset of outputs at redshifts, $z= 0$, $0.5$, $1.0$ and $2.0$.  
Unless otherwise noted, all of our analysis is for primary halos over the total mass limit of  $5 \times 10^{13} \hinv \msun$.

%%%%%%%%%%%%%%%%%%%%%%%%%%%%%%%%%%%%%%%%%%%%%%%%%%%%%%%%%%%%%%%%%%%%%%%%%
\subsection{Bulk Halo Properties} \label{sec:bulkmeas}

With the primary halo samples identified, we calculate bulk properties for them that we roughly classify into ``structural'' and ``observable'' categories.  The former includes dark matter velocity dispersion, ICM mass fraction, gas mass-weighted temperature, and halo concentration while the latter includes X-ray luminosity and spectroscopic-like temperature, thermal Sunyaev-Zel'dovich effect, and a dimensionless ICM emission measure.

As the MGS simulations are SPH treatments of the gas, integrals over volume map 
to summations over all particles, via $\int dV \rho^n \rightarrow \Sigma_i m_i \rho_i^{n-1}$.   
We consider two measures of ICM temperature.  First, we consider 
the mass-weighted temperature.  As GADGET-2 is a Lagrangian simulation with equal mass 
gas particles, the mass-weighted temperature is simply the average temperature of the 
particles in the halo:
\begin{equation}
T_m = \frac{1}{M} \int_V dV \rho \, T \rightarrow \frac{1}{N} \sum_i^N T_i .
\end{equation}

We also calculate the spectroscopic-like temperature, $\tsl$, as defined in 
\cite{mazzotta:04}, 
\begin{equation}
\tsl = \frac{\int n^2 T^{\alpha - 1/2} dV}{\int n^2 T^{\alpha-3/2} dV} \rightarrow 
\frac{\sum_i^N \rho_i T_i^{\alpha - 1/2}}{\sum_i^N \rho_i T_i^{\alpha-3/2}},
\end{equation}
with $\alpha = 0.75$.  The spectroscopic-like temperature offers a good match to the temperature 
derived from a one-component fit to an X-ray spectrum, but is far simpler to compute.  

We calculate X-ray luminosities, 
% assuming that the emission is bremsstrahlung, such that 
% the cooling function $\Lambda(T) \propto T^{1/2}$, 
\begin{equation}
L = \int_V dV \rho^2 \Lambda(T) \rightarrow \sum_i^N \rho_i \tilde{\Lambda}(T_i) ,
\end{equation}
using MEKAL tables assuming fixed $0.3$ solar metallicity to calculate $\tilde{\Lambda}(T)$. 
We do this in energy bands of  0.7-2.0 keV, 0.7-5.0 keV, and 
0.7-7.0 keV, and also compute a bolometric luminosity, $\elbol$ using a wide photon energy range of  [0.1-40.0] keV.   
These tables include both continuum and emission lines derived from an assumption of collisional ionization equilibrium \citep{mewe:03}.  

We also calculate the global thermal Sunyaev-Zel'dovich signal, $Y$, parameter, following the convention 
% which is an X-ray measurement that is an analog to the Sunyaev-Zeldovich (SZ) signal, 
presented in \cite{dasilva:00,springel:01,kravtsov:06}.  
\begin{equation}
Y = \left(\frac{k_B \sigma_T}{m_ec^2} \frac{1}{A}\right) \int_V dV n_e T_e \rightarrow \left(\frac{k_B \sigma_T}{m_ec^2} \frac{1}{A} \frac{M_{gas}}{m_p}\right) \sum_i^N T_i.
\end{equation}
With $k_B$ Boltzmann's constant, $\sigma_T$ the Thomson cross-section, $m_e$ the electron mass, and $m_p$ the proton mass, and $A$ the halo comoving area.    The above expression yields $Y$ in units of $h^{-2} \mpc^2$.

%%%%%%%%%%%%%%%%%%%%%%%%%%%%%%%%%%%%%%%%%%%%%%%%%%%%%%%%%%%%%%%%%%%%
\subsection{Radial Profile Measures}\label{sec:profile}

In addition to bulk cluster properties, we include two measures of radial 
structure in our analysis: halo concentration of the total mass and a dimensionless emission measure for the hot gas.  

We measure the halo concentration, $c$, after fitting 
an NFW density profile \citep{nfw:97}, 
\begin{equation}
\frac{\rho(r)}{\rho_{crit}} = \frac{\delta_c}{(r/r_s)(1 + r/r_s)^2},
\end{equation}
to the radially-binned, total mass density profile of each halo.   Following NFW, we take the concentration, $c$, to be defined at a density threshold of $200$ times the critical density, $c = r_{200}/r_s$.  
% We find the halo concentration via
% \begin{equation}
% \delta_c = \frac{200}{3}\frac{c^3}{[\ln (1+c) - c/(1+c)]}.
% \end{equation}
Halo concentration is a good proxy for halo formation epoch \citep{wechsler:02, busha:07}.  We show below that baryon loss in the \PH case has a non-neglgible effect on halo concentrations.  

We also measure a dimensionless emission measure, or clumping factor, $Q$, 
 that measures the contribution of gas density structure to the halo luminosity.  
 Using a scaled radius, $y = r/r_{200}$, we write the radial gas density profile of a given halo as
\begin{equation}
\rho(yr_{200}) = \ficm (200 \rho_c) g(y),
\end{equation}
where $\ficm \equiv M_{\rm gas}(<r_{200})/M_{200}$ is the halo's ICM mass fraction within $r_{200}$.   With the overall ICM mass fraction factored out, $g(y)$ becomes a dimensionless structure function normalized by $3 \int_0^1dyy^2g(y)=1$.   

The second moment of $g(y)$ defines the dimensionless emission measure 
\begin{equation}
\hat{Q}(T) = (3/4\pi) \int {\rm d^3} yg^2(y).  
\end{equation}
This definition allows the X-ray luminosity scaling to be written as 
\begin{equation}
L \propto \rho_c M_{200} \Lambda(T) \ \ficm^2  \  \hat{Q}.
\end{equation}
We show below how lower values of both $\ficm$ and $\hat{Q}$ for the \PH case drive X-ray luminosities down by an order of magnitude at $10^{14} \hinv\msol$ relative to the \GO treatment.

%%%%%%%%%%%%%%%%%%%%%%%%%%%%%%%%%%%%%%%%%%%%%%%%%%%%%%%%%%%%%%%%%%%%%
%%%%%%%%%%%%%%%%%%%%%%%%%%%%%%%%%%%%%%%%%%%%%%%%%%%%%%%%%%%%%%%%%%%%%
\section{Mean Scaling Relations}\label{sec:scaling}

\begin{figure*}
\begin{center}
$\begin{array}{cc}
\includegraphics[width=2.25in]{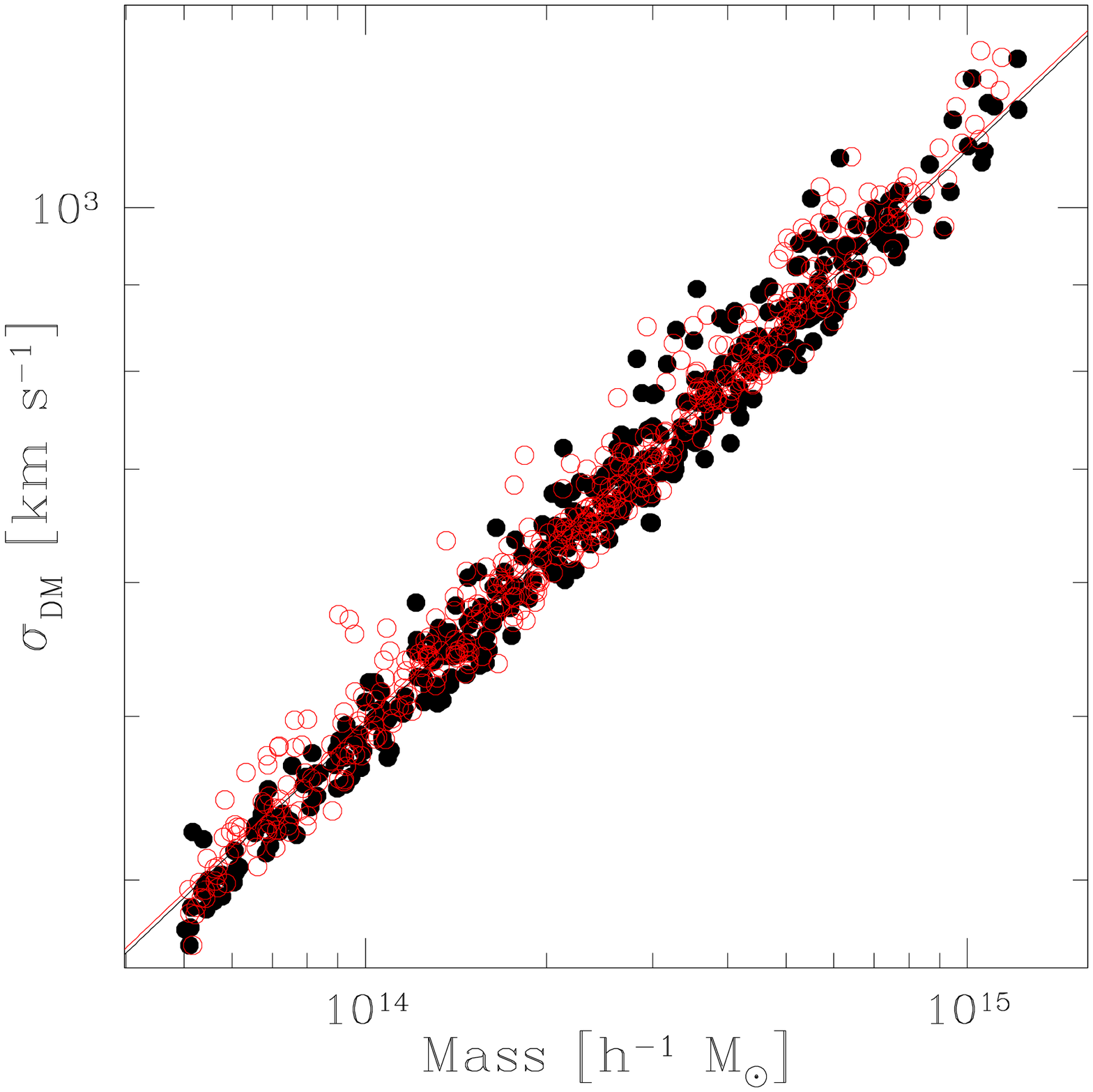} &
\includegraphics[width=2.25in]{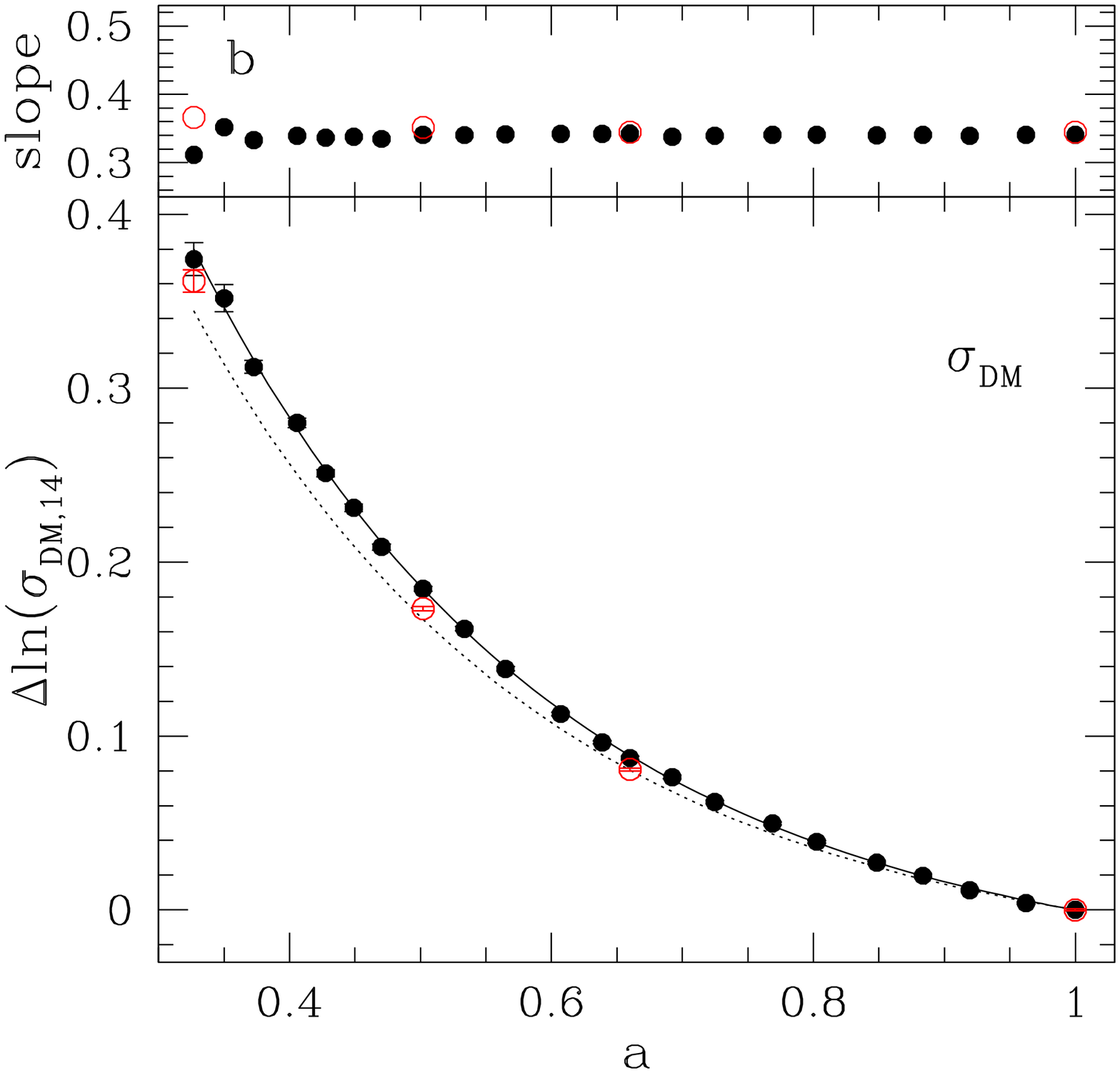} \\
\includegraphics[width=2.25in]{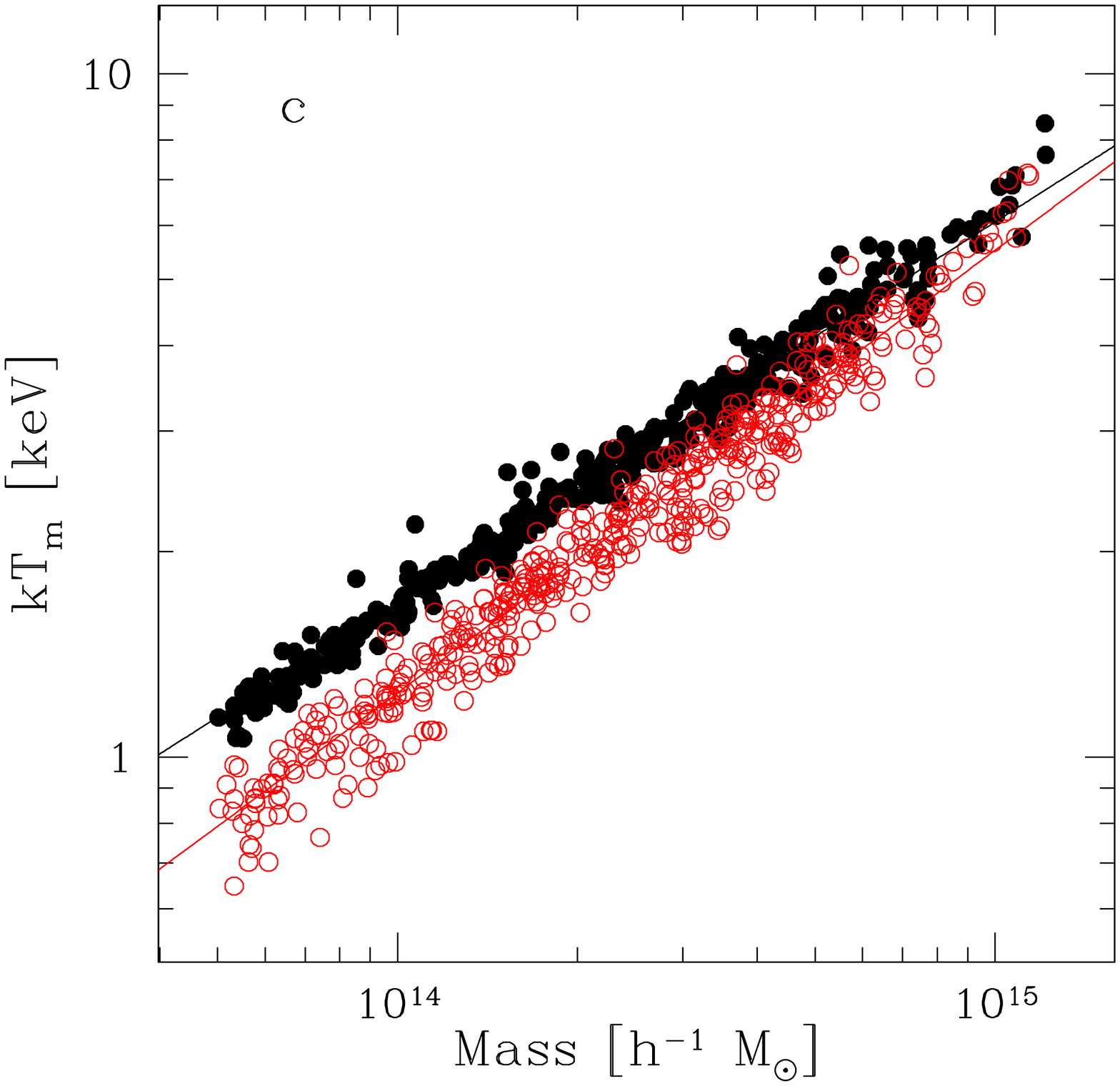} &
\includegraphics[width=2.25in]{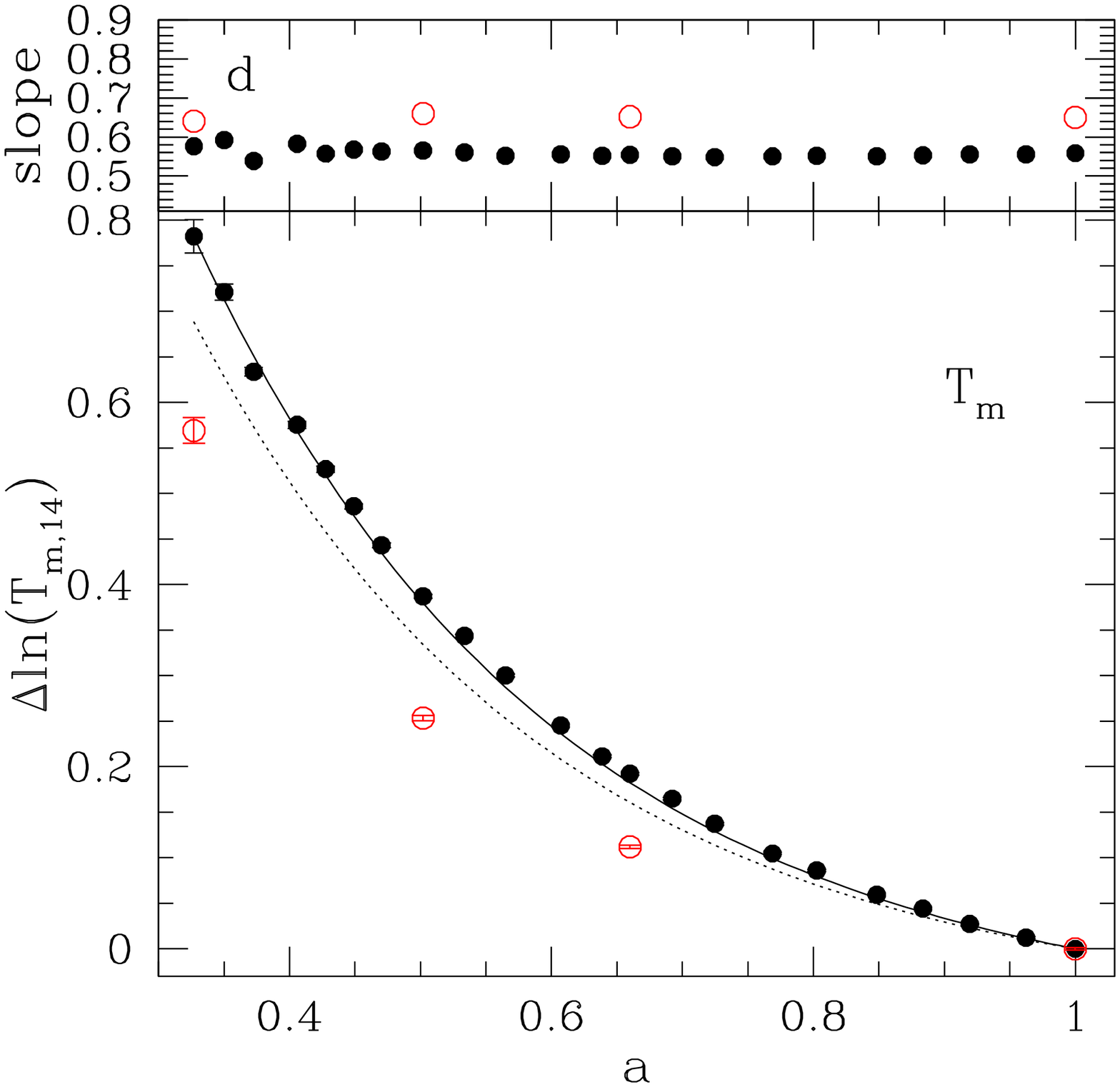} \\
\includegraphics[width=2.25in]{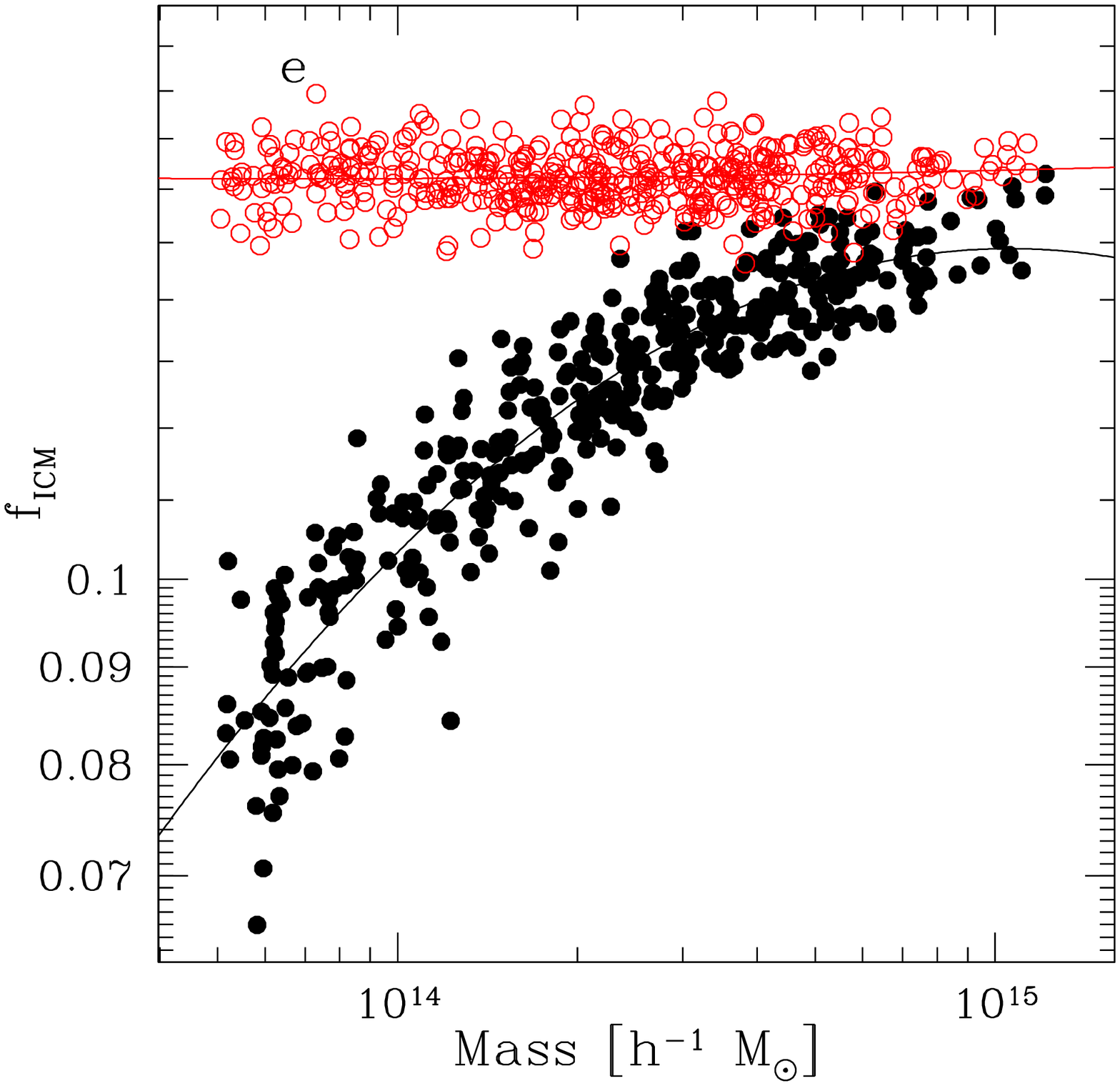} &
\includegraphics[width=2.25in]{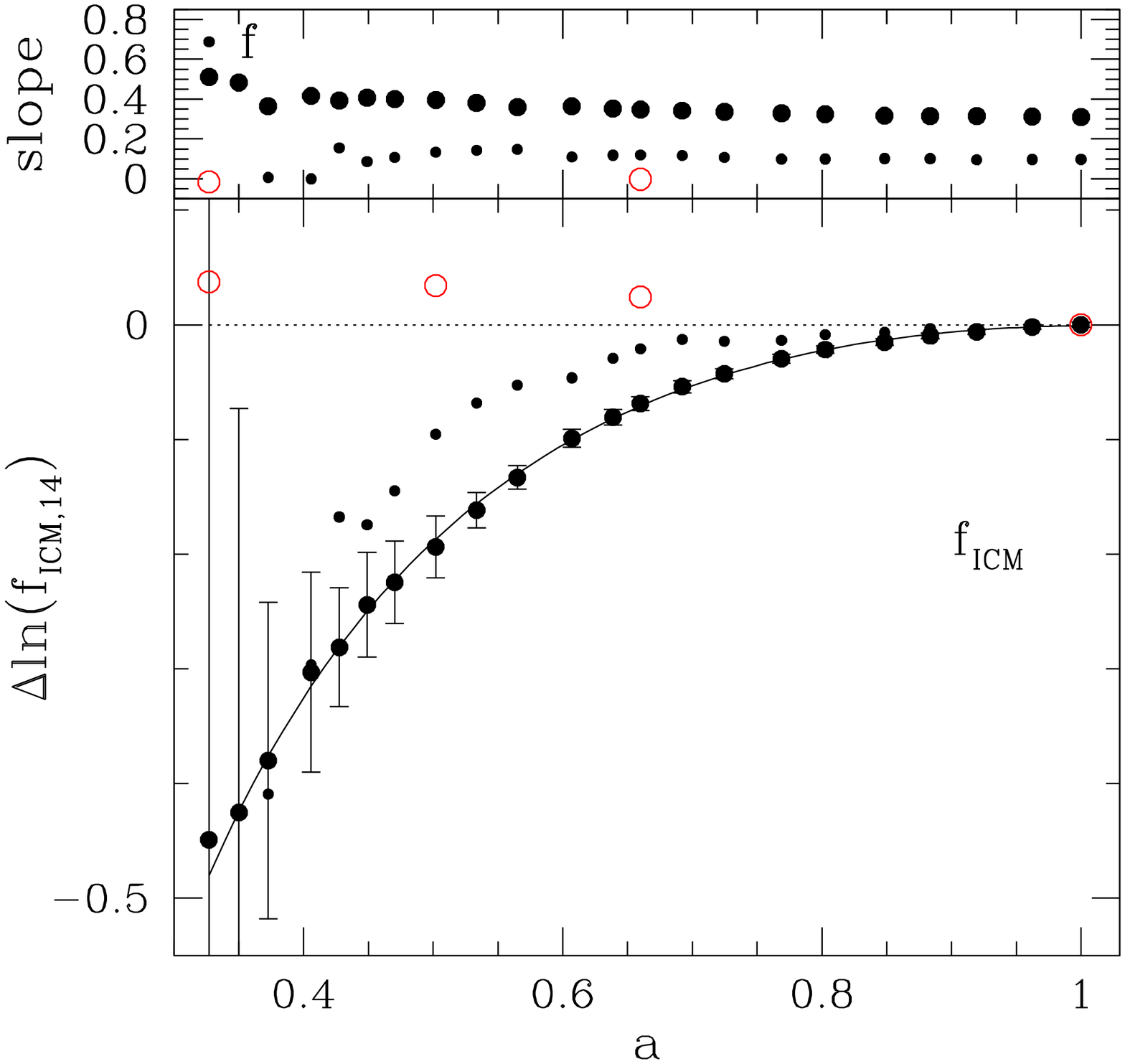} \\
\includegraphics[width=2.25in]{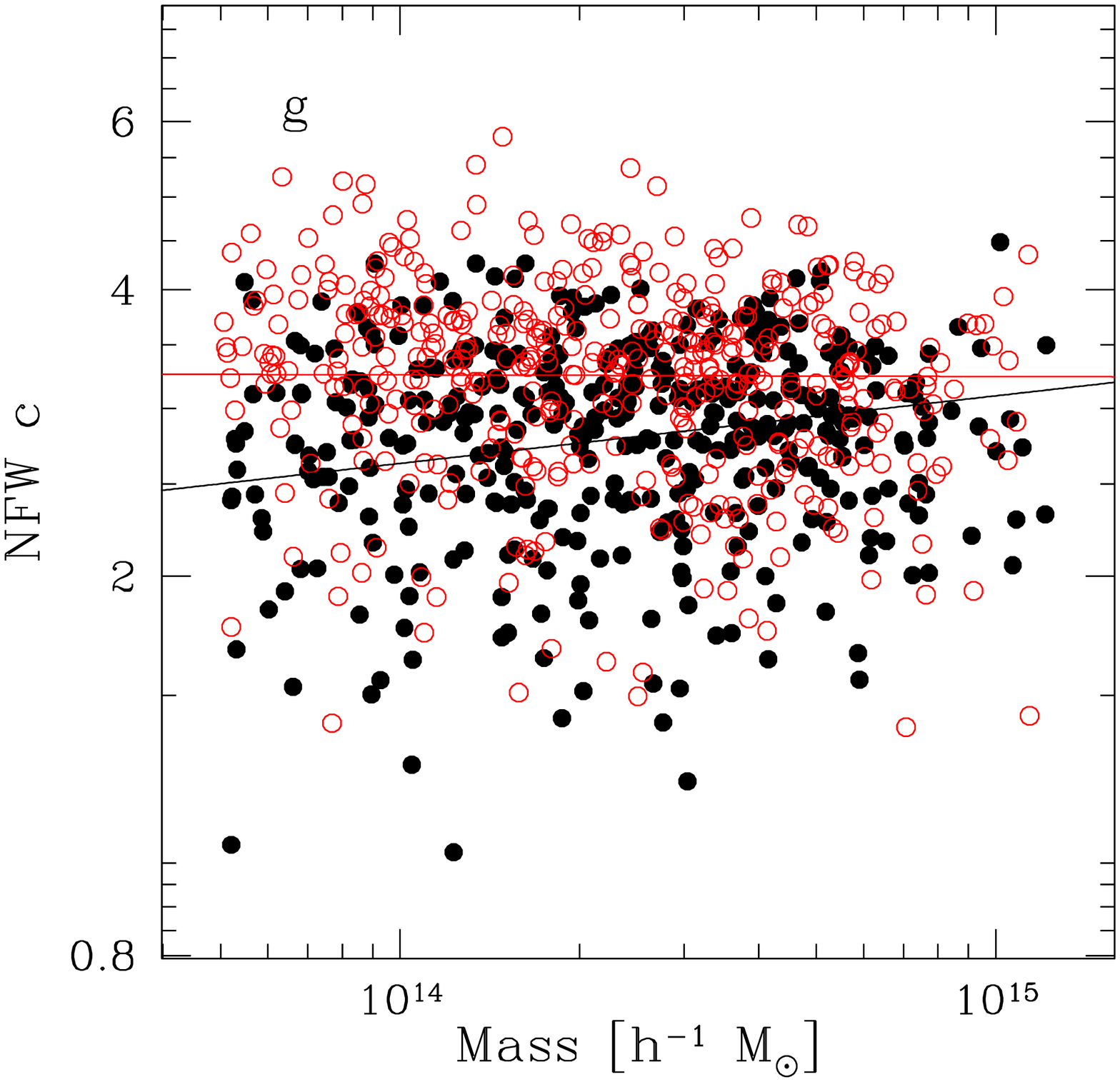} &
\includegraphics[width=2.25in]{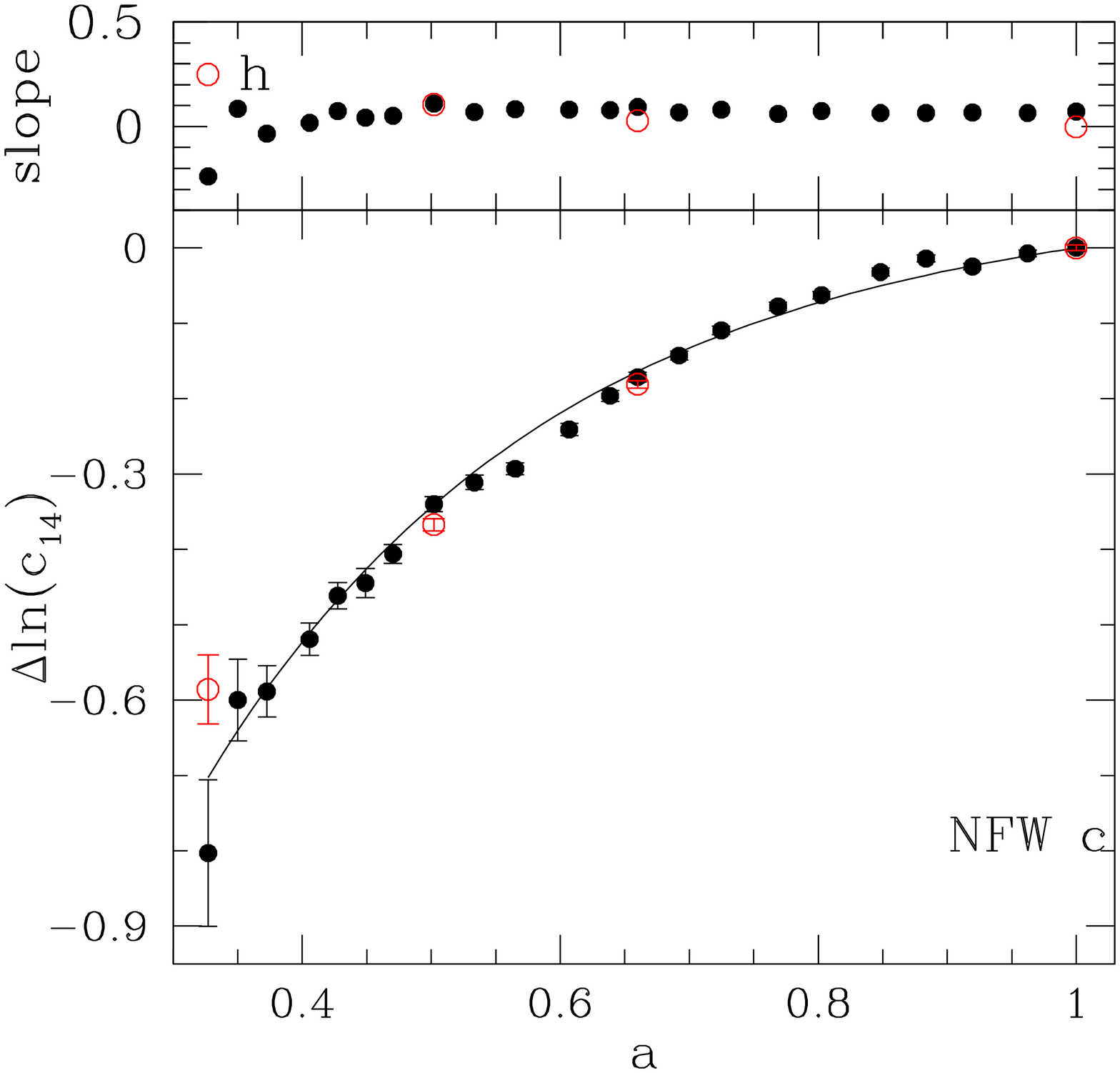}
\end{array}$
\end{center}
\caption{Scaling relations at redshift zero (left) and redshift evolution of fit parameters (right) for $\sigmadm$, 
$T_m$, $\ficm$, and the NFW concentration $c$ (top to bottom).   The \PH model data are black, filled points and 
the \GO are red, open points.  Subsamples of halos are shown in the left panels, drawn in a manner that shows $\sim 10\%$ of the overall population with nearly uniform mass sampling.   Solid lines show the best-fit scalings derived from the entire sample.   
The right-hand panels present the evolution of the slope and the shifts in normalization, equation~(\ref{eqn:dels14}).  
Solid lines show fits to log-linear behavior, equation~(\ref{eqn:s14fit}), except $ficm$ which is quadratic in $\ln M$, while the dotted lines give the self-similar predictions listed in Table \ref{tab:evol}.  
% For the evolution of the \PH parameters, we plot a subset of time outputs evenly spaced in $a$.
\label{fig:scalestructure}}
\end{figure*}

To characterize the mean behavior of halo properties, we perform  
a linear least squares fit to the natural log of the $i^{\rm th}$ signal as a function 
of mass and redshift, using the form 
\begin{equation} 
\langle s_i \rangle(\mu,a)  = s_{i,14}(a) + \alpha_i(a) \mu,   
\label{eqn:lsqfit}
\end{equation} 
where $s_i \equiv \ln S_i$ and $\mu \equiv \ln ( M/\hinv 10^{14} M_\odot)$.  Brackets represent averaging in narrow mass bins at fixed epoch, so $\alpha_i(a)$ is the slope and $s_{i,14}(a)$ the normalization at  $10^{14} \hinv M_\odot$ of the $i^{\rm th}$ signal at redshift, $z = a^{-1}-1$.   In the \PH case, one expects the entropy floor to introduce curvature into the scaling relations \citep{voit:02}, and so we extend this model to a quadratic with respect to $\mu$ for the ICM mass fraction and related measures for this case. 

At each redshift, the mass scalings are derived from the sample of primary  halos with $M > 5 \times 10^{13} \hinv \msun$.  We then fit the evolution of the normalization, presented as a shift relative to the present value, 
\begin{equation} 
\Delta s_{i,14}(a) = s_{i,14}(a) - s_{i,14}(1)  , 
\label{eqn:dels14}
\end{equation} 
to the form 
\begin{equation} 
\Delta s_{i,14}(a) =  \beta \ln(E(a)),
\label{eqn:s14fit}
\end{equation} 
where $E(a) = H(a)/H_0$.   For the \PH model, this form is a poor fit to $\ficm$-related quantities.  We find a better fit in terms of a quadratic in $\ln\,a$, 
\begin{equation} 
\Delta s_{i,14}(a) = \epsilon_0 + \epsilon_1 \ln(a)  + \epsilon_2 (\ln(a))^2,
\label{eqn:s14lnafit}
\end{equation} 

In this section, we present mean scaling relations and discuss deviations from self-similar expectations \citep{kaiser:86} for structural (\S\ref{sec:structure}), and observable (\S\ref{sec:xraybulk}) properties.  We then demonstrate good agreement between X-ray observations and the  low-redshift PH model results.  
Deviations from mean scaling behavior, in the form of a signal covariance matrix, $\langle (s_i- \langle s_i\rangle) (s_j - \langle s_j \rangle) \rangle$, are presented in \S\ref{sec:covar}.

%%%%%%%%%%%%%%%%%%%%%%%%%%%%%%%%%%%%%%%%%%%%%%%%%%%%%%%%%%%%%%%%%%%%%%%%%
\subsection{Structural Quantities}\label{sec:structure}

The left panels of Figure~\ref{fig:scalestructure} present scaling relations as a function of mass at $z=0$ for sub-samples of the \GO and \PH halo samples.  Four structural measures are presented:  dark matter velocity dispersion, $\sigmadm$; intracluster gas mass fraction, $\ficm$; mass-weighted gas temperature, $T_m$; and NFW concentration, $c$.   Best-fit parameters to the mass scaling, equation~(\ref{eqn:lsqfit}), are presented in Table ~\ref{tab:params}  The right panels show redshift evolution of the slopes and the shifts in normalization, equation~(\ref{eqn:dels14}), for each signal.  Error bars in the fit parameters are derived from bootstrapping resampling of the samples.  In general, the uncertainties on the best fit parameters are very small,  $\sim 0.1\%$ at redshift $z=0$, and are much smaller than the intrinsic scatter at fixed mass about the median power law relation.  The errors grow larger at higher redshifts; the mass-limited sample size drops below 100 at  $z=1.8$ for the \PH case.   Fits to the redshift evolution of the normalization, equation~(\ref{eqn:s14fit}) are presented in Table ~\ref{tab:evol}.

{\it Dark matter velocity dispersion.}  The velocity dispersion of the dark matter particles is a 
fundamental measure of the virial state of a halo.  Observationally, the galaxy velocity dispersion tracks ICM temperature in a manner consistent with virial expectations, but the possibility of a $\sim 10\%$ bias relative the dark matter is still allowed  \citep{biviano:04,  becker:07}.  In both simulations,  Figure~\ref{fig:scalestructure}a  shows that the scaling with mass 
is slightly shallower than the self-similar prediction, $\sigmadm \propto M^{1/3}$.  The dark matter velocity dispersion at $10^{15} \hinv\msol$ is in good agreement with the $1082.9 \pm 4.0 \kms$ value derived from a suite of N-body simulations by \citet{evrard:08}.  As discussed in \citet{evrard:08}, a slight suppression of the slope is expected from finite particle resolution when the low mass halo cutoff corresponds to a few thousand particles, as is the case here.   The fact that the PH and GO slopes are similar ($0.341$ and $0.345$) indicates that the difference in gas physics treatments is not responsible for the shift.  
In both \PH and \GO cases, the slope remains constant to high redshift, as seen in the upper right panel of Figure ~\ref{fig:scalestructure}.  The evolution of the normalization, $\sigma_{DM,14}$ shown in Figure~\ref{fig:scalestructure}b, 
% agrees well between the two simulations:  we find 
% $\sigma_{DM,14} = 496.21 \kms \pm 0.29$ in the \GO simulation and $\sigma_{DM,14} = 494.37 \pm 0.31 
% \kms$ in 
% the \PH simulation. This 
% is in good agreement with the result presented in \cite{evrard:08}, which has a normalization of 
% $\sigma_{DM,14} = 500.2 \pm 6.0 \kms$. The evolution of the normalization 
deviates slightly from the self-similar prediction, $\sigmadm \propto [E(z) M]^{1/3}$, with the velocity dispersion being $1-2\%$ higher at $z=1$ (values of $\beta$ are given in Table ~\ref{tab:evol}).  Despite strong baryon content differences discussed below, the dark matter virial scaling under the \PH and \GO treatments remain remarkably consistent.   

\begin{deluxetable*}{lcc|cc}
\tablecaption{Redshift Zero Mass Scalings \tablenotemark{a} \label{tab:params}}
\tablehead{
\colhead{} & \multicolumn{2}{c}{\PH Simulation} & \multicolumn{2}{c}{\GO Simulation} \\
\colhead{Signal}  & \colhead{$s_{14}$} & \colhead{$\alpha$} & \colhead{$s_{14}$} & \colhead{$\alpha$} }
\startdata
$\sigmadm$\tablenotemark{b} & $6.1990 \pm 0.0006$ & $0.341 \pm 0.001$ & $6.2037 \pm 0.0006$  & $0.345 \pm 0.001$  \\
$kT_m$\tablenotemark{b}  & $0.5121 \pm 0.0008$ & $0.559 \pm 0.002$ & $0.209 \pm 0.001$ & $0.650 \pm 0.002$ \\
$k\tsl$\tablenotemark{b} & $0.605 \pm 0.001$ & $0.576 \pm 0.002$ & $0.163 \pm 0.003$ & $0.576 \pm 0.005$ \\
$Y$\tablenotemark{b} &$-12.810 \pm 0.002$ & $1.825 \pm 0.003$ & $-12.642 \pm 0.002$ & $1.651 \pm 0.003$ \\
$\elbol$\tablenotemark{b} & $-1.653 \pm 0.003$ & $1.868 \pm 0.006$ &  $0.622 \pm 0.005$ & $1.079 \pm 0.008$ \\
NFW $c$ &$ 0.966 \pm 0.005$ & $0.071 \pm 0.007$ & $1.181 \pm 0.004$ & $-0.002 \pm 0.006$  \\
$\hat{Q}$ & $0.461 \pm 0.002$ & $0.161 \pm 0.003$ &  $1.219 \pm 0.001$ & $0.030 \pm 0.003$  \\
\enddata
\tablenotetext{a}{Fit Parameters to equation~(\ref{eqn:lsqfit}) with uncertainties from Monte Carlo re-sampling.}
\tablenotetext{b}{Units: $\sigmadm$ ($\kms$);  $kT_m$ or $k\tsl$ (keV); $\elbol$ ($10^{44} \cgsflux$). }
\end{deluxetable*}

{\it Mass-weighted temperature.} The mass-weighed temperature, $T_m$, a useful probe of the hydrodynamic state 
of the ICM, is known to have small scatter with respect to mass in simulations \citep{evrard:96,bryan:98,borgani:04}.  Figure~\ref{fig:scalestructure}c  shows that the slope of the \GO scaling relation agrees with the self-similar scaling, $T \sim M^{2/3}$, at the few percent level, whereas the \PH slope is significantly less steep, $\alpha = 0.559 \pm 0.002$.  As all gas particles receive the same entropy boost at $z \sim 4$ in \PH, the resultant fractional increase in the characteristic entropy of a halo is larger for low-mass halos than for high-mass halos. The elevated initial entropy drives the tilt as well as a $\sim35\%$ higher normalization of the $T_m-M$ relation in the \PH simulation.  But the effect diminishes at higher masses; the \PH halos at $10^{15} \hinv \msol$ are only $10\%$ hotter than their \GO counterparts. 

The slopes of the $T_m-M$ relation do not evolve strongly with redshift in either model.  The normalizations, however, 
do not follow the self-similar expectation, with the \GO simulation lying $\sim 10\%$ low and the \PH case $5\%$ high at $z = 1$ relative to $T_m \sim E(a)^{2/3}$ scaling \citep{kaiser:86}.  The self-similar prediction 
is for halos in perfect hydrostatic equilibrium, and mergers are known to drive deviations from hydrostatic equilibrium.  However, the total kinetic energy of the gas --- the sum of thermal plus bulk kinetic, or turbulent components --- should scale more closely to self-similar expectations.  

\begin{figure}
\plotone{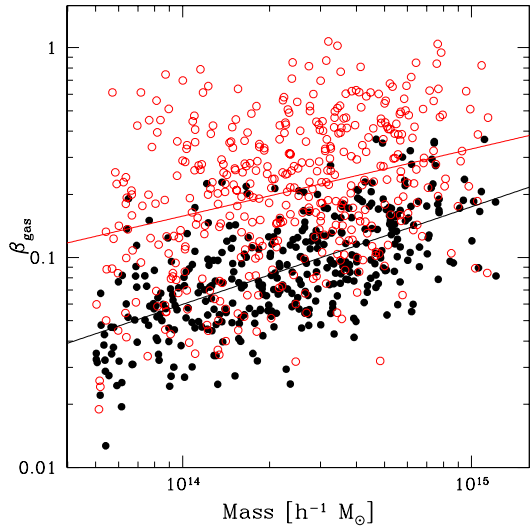}
\caption{The normalized bulk kinetic energy in gas turbulent motions, $\betagas = \sigmagas^2 / (kT_m/\mu m_p)$, is shown as a function of mass for sub-samples of the \PH (filled, black points) and \GO (open, red points) halos.   Lines show best-fit mean scalings.    
\medskip
\label{fig:betagas}}
\end{figure}

We define a dimensionless measure of the kinetic energy content of the gas, $\betagas = \sigmagas^2 / (kT_m/\mu m_p)$, where $\mu$ is the mean molecular weight of the gas.    
Figure~\ref{fig:betagas}. shows values of $\betagas$ for a subsample of halos at $z=0$.   
The values differ strongly in the two simulations, with mean values at $10^{14} \hinv \msun$ of $0.16$ in the \GO simulation and $0.06$ in the \PH model.   The positive scaling with mass reflects the later formation epoch of high-mass halos.  
The next generation of X-ray telescopes, such as IXO\footnote{http://ixo.gsfc.nasa.gov/}, will have high-resolution spectroscopy capable of measuring  $\sigmagas$, and thus be able to discriminate between the model predictions in Figure~\ref{fig:betagas}. 

\begin{figure}
\plotone{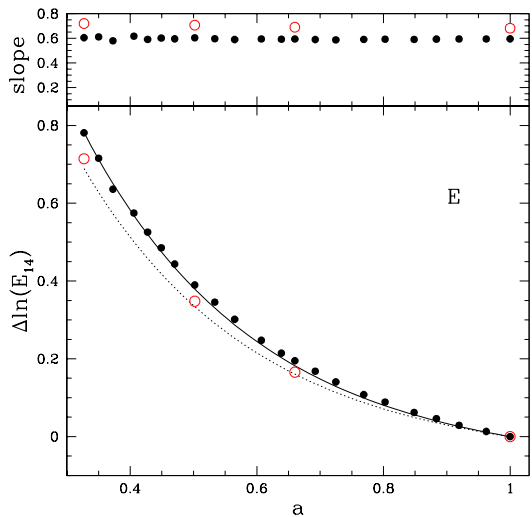}
\caption{The evolution of the normalization and slope of the total kinetic energy of the gas in the 
\PH  (filled, black points) and the \GO  (open, red points) simulations.  The dotted 
line is the self-similar prediction and the solid line is 
the measured evolution in the \PH simulation.   
% For the evolution 
% of the \PH energy, we plot a subset of time outputs evenly spaced in $a$.
\label{fig:evole}}
\end{figure}

We measure the total gas kinetic energy, 
\begin{equation}
E_{\rm tot} = \frac{kT_m}{\mu m_p} + \sigmagas^2, 
\end{equation}
and fit this to the standard form, equation~(\ref{eqn:lsqfit}).  
% Even if the mass-weighted temperature does not evolve in a self-similar fashion, we expect 
% the total gas energy to evolve self-similarly in an adiabatic simulation.  Indeed, we see in 
Figure ~\ref{fig:evole} shows the redshift evolution of the total gas kinetic energy in both models.   The \GO simulation respects the self-similar scaling, $E_{\rm tot} \sim [M E(a)]^{2/3}$ at the few percent level to redshifts $z=1$.  At a given mass, the velocity dispersion contribution to the total kinetic energy increases with redshift, and this enhanced turbulence is responsible for driving the mass-weighted temperature away from self-similar scaling, as seen in Figure~\ref{fig:scalestructure}.  For the \PH case, the degree of turbulence is much lower, and the total energy remains higher at high redshift due to the influence of the initial entropy injection.  

{\it Baryon fraction.}    Figure \ref{fig:scalestructure}e,f shows the scaling of the hot gas fraction, $\ficm$, with mass at $z=0$ along with the redshift evolution of the slope and normalization.  For reasons discussed below, we show evolution in the \PH normalization and slope at a mass of $5 \times 10^{14} \hinv \msun$ as well as the fiducial mass of $10^{14} \hinv\msun$.  

% Although $\ficm$ is difficult to measure observationally, as it requires measuring the cluster mass, the differences in cluster 
% properties in \PH and \GO are driven by the different states of the ICM.  
At redshift zero, the distribution of $\ficm$ values within $r_{200}$ in the \GO simulation is simple: halos have a slightly depleted 
baryon fraction, with mean $\ficm = 0.90$ , and a dispersion of $0.04$.  There is no trend with mass, 
as the slope is $0.001 \pm 0.019$, a result consistent with previous SPH 
simulations done at similar resolution and evaluated at 
$r_{200}$  by \cite{crain:07} and \cite{ettori:06}.   The simulation of the MareNostrum 
universe \cite{gottlober:07} has twice our mass resolution, and a baryon fraction of $\ficm = 0.92$ 
at a virial radius, $r_{vir}$, that encompasses a mean density of $\sim 100$ times the critical density.  This small but significant increase is consistent with the radial trend seen in our simulation.  
Our value of $\ficm$ is lower than the $0.97$ value measured by \cite{kravtsov:05} at $r_{vir}$ 
 in their adiabatic AMR (Adaptive Mesh Refinement) code.   That study included a comparison of SPH and AMR simulations evolved under gravity only, and they note a statistically-significant offset of $\sim 5\%$, with the AMR simulation having higher baryon fraction.

\begin{deluxetable}{c|ccc}
\tablecaption{Redshift Zero \PH Quadratic Mass Scalings  \tablenotemark{a} \label{tab:ficm}}
\tablehead{ \colhead{Signal} & \colhead{$s_{14}$} & \colhead{$\alpha_1$} & \colhead{$\alpha_2$} 
}
\startdata
$\ficm$ & $-2.270 \pm 0.002$ & $0.310 \pm 0.009$ & $-0.0661 \pm 0.0061$ \\
$Y$\tablenotemark{b} & $-12.790 \pm 0.005$ & $1.864 \pm 0.018$ & $-0.0586 \pm 0.0017$ \\
$\elbol$\tablenotemark{b} & $-1.641 \pm 0.012$ & $ 1.892 \pm 0.043$ & $-0.0351 \pm 0.0042$ \\
\enddata
\tablenotetext{a}{Fits of the \PH signal-mass relations at $z=0$ to a quadratic, $\ln S = s_{14} + \alpha_1 \ln M + \alpha_2 (\ln M)^2$, with mass in units of $10^{14} \hinv\msun$. }
\tablenotetext{b}{Units: $Y$ ($h^{-2} \mpc^2$); $\elbol$ ($10^{44} \cgsflux$). }
\end{deluxetable}

As has been shown in previous simulations \citep{bialek:01, borgani:01, muanwong:06, younger:07}, preheating has a dramatic effect on the hot gas fraction of massive halos.   Low-mass halos can lose half of their baryons within $r_{200}$ while the most massive halos are depleted by only $\sim 10\%$ relative to the \GO case.   A power-law form is a poor representation of the \PH mean behavior of $\ficm$ with mass, so we extend the model to a quadratic in $\ln M$ and give best-fit parameters in Table~\ref{tab:ficm}.   We present a comparison with observed gas fractions in \S\ref{sec:meanobs} below. 

\begin{deluxetable}{lcc}
\tablecaption{\PH Normalization Evolution in $\ln(E(a))$ \tablenotemark{a} \label{tab:evol}}
\tablehead{
\colhead{Signal} & \colhead{$\beta$} & \colhead{Self-Similar} }
\startdata
$\sigmadm$ & 0.34 & $1/3$ \\
$T_m$ & 0.76 & $2/3$ \\
$\tsl$ & 0.73 & $2/3$ \\
$\elbol$ & 1.39 & $7/3$ \\
$\ficm$ & -0.44 & 0 \\
$Y$ & 0.33 & $2/3$ \\
$\hat{Q}$ & -0.13 & N/A \\
$c$ & -0.68 & N/A \\
\enddata
\tablenotetext{a}{Equation~\ref{eqn:s14fit}.}
\end{deluxetable}

\begin{deluxetable}{lccc}
\tablecaption{\PH Normalization Evolution in $\ln(a)$ \tablenotemark{a} \label{tab:quadevol}}
\tablehead{
\colhead{Signal} & \colhead{$\epsilon_0$} & \colhead{$\epsilon_1$} & \colhead{$\epsilon_2$} }
\startdata
$\ficm$ & $-7.64 \times 10^{-4}$  & 0.0143 &  -0.371 \\
$Y$ & $-1.49 \times 10^{-3}$ & -0.308 & -0.0141 \\
$\elbol$ & $-9.44 \times 10^{-4}$ & -0.786 & 0.477 \\
\enddata
\tablenotetext{a}{Equation~\ref{eqn:s14lnafit}.}
\end{deluxetable}

In the \GO simulation, the slope of the gas fraction remains consistent with zero at all redshifts, while the mean gas fraction increases slightly, rising from $\ficm = 0.90$ at $z=0$ to $0.93$ at $z = 1$.  The latter effect is consistent with the findings of \cite{gottlober:07} in the Marenostrum simulation, where the mean baryon fraction increases from $0.92$ at $z = 0$ to $\ficm = 0.94$ at $z=1$.  While the mechanism responsible for this slight drift is not fully understood, energy transfer from the dark matter to the gas during mergers may play a role \citep{pearce:94,mccarthy:07}.   

The redshift evolution of the baryon fraction in the \PH simulation depends on mass scale.  Gas expelled from low-mass halos by  
preheating at $z \sim 4$ can accrete back onto descendant halos at later stages in the merger hierarchy, in a manner that depends on the characteristic entropy of the later halo.  At the fiducial $10^{14} \hinv\msun$ normalization scale, halos at $z=1$ have a $\sim 20\%$ lower gas fraction compared to $z=0$, and the local slope of the $\ficm - M$ relation is also steeper at higher redshift.   The effects of preheating dominate over the universal expansion at the $10^{14} \hinv \msun$ mass 
scale; hence the evolution of the normalization cannot be simply described as a power of $E(z)$.  For the baryon fraction, and for other halo parameters which 
depend strongly on the baryon fraction, we fit the evolution as a quadratic in $\ln (a)$, with the best-fit parameters presented in Table \ref{tab:quadevol}.

At a higher mass scale of $5 \times 10^{14} \hinv\msun$, the effects are milder, and the local slope is very close to the slope of the \GO model.  The most massive halos are less affected by preheating, and their baryon fractions are therefore more appropriate to use as indicators of cosmology \citep{pen:03,allen:04,allen:08}.   We note that the mean gas fraction shift at $5 \times 10^{14} \hinv\msun$ in the \PH model is comparable to the 20 percent uniform prior on gas fraction applied by \cite{allen:08} in their analysis of a {\sl Chandra} sample of 42 clusters with $kT > 5 \kev$ extending to $z=1.1$.  The cosmological constraints from that work would thus not be strongly affected if a \PH model prior on $\ficm$ behavior were imposed.  

{\it NFW concentration.}   Figure~\ref{fig:scalestructure}g,h show the behavior of the NFW concentration, $c$, derived from fits to the total mass density profile of each halo.  At $10^{14} \hinv\msun$, the mean concentration in the \GO simulation, $c = 3.3$, is substantially lower than the value of $5.2$ found for the original Millennium Simulation with only dark matter \citep{neto:07,gao:08}.   The mean concentration in the \PH simulation, $c = 2.6$, is lower than the \GO value.   As discussed above, the \PH halos have a lower baryon fraction than \GO halos, and the increasing baryon loss at low masses has the effect of tilting the $c-M$ relation so that the slope is positive rather than weakly negative.

%%%%%%%%%%%%%%%%%%%%%%%%%%%%%%%%%%%%%%%%%%%%%%%%%%%%%%%%%%%%%%%%%%%%%%%%%%%%%%%%%%%%%
\subsection{X-ray and SZ Signals}\label{sec:xraybulk}

We now turn to common bulk observed properties of clusters: the SZ decrement, $Y$, and X-ray spectroscopic-like temperature, $\tsl$, luminosity, $\elbol$, and the emission measure $\hat{Q}$. 

\begin{figure*}
\begin{center}
$\begin{array}{cc}
\includegraphics[width=2.25in]{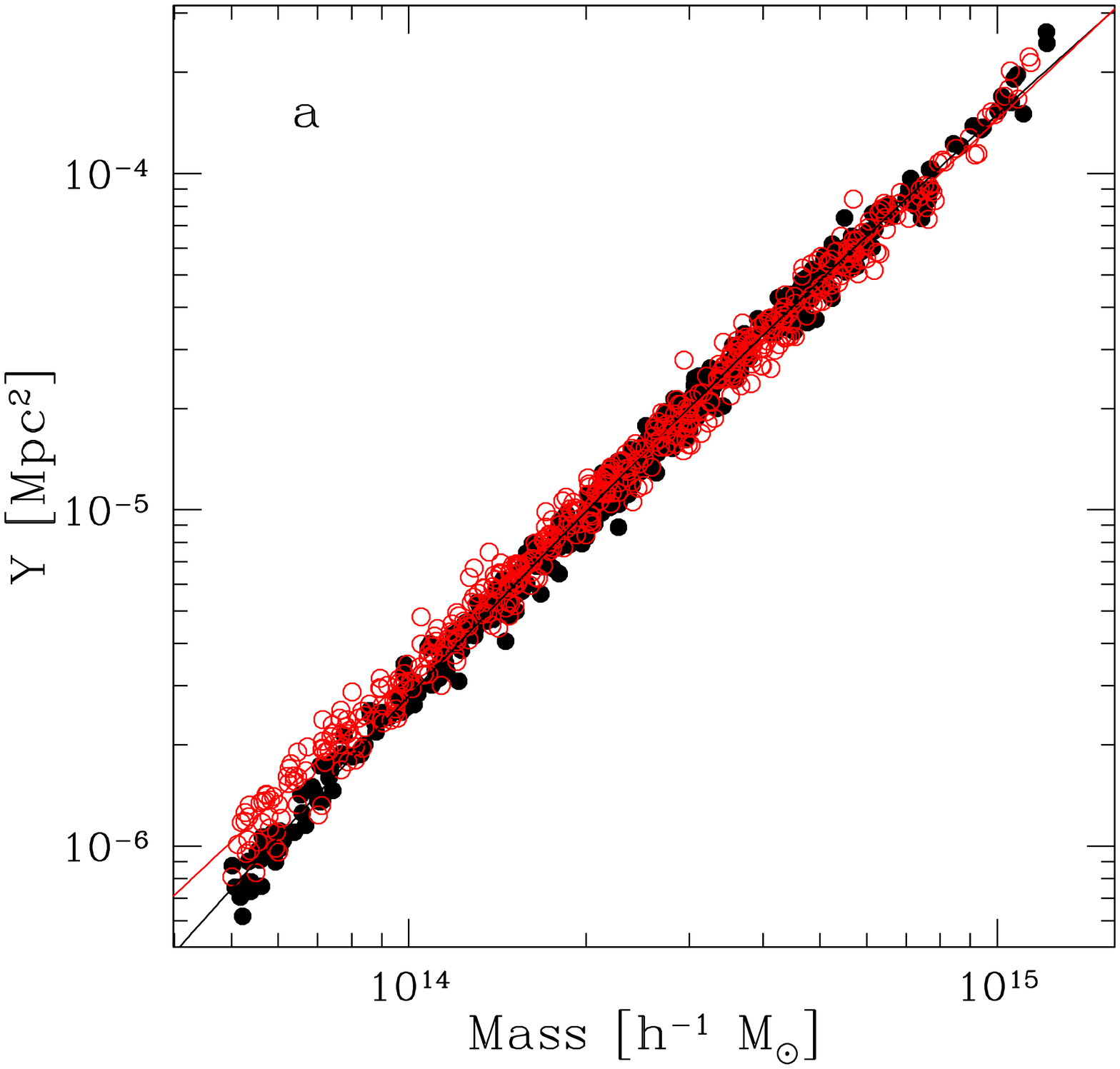} &
\includegraphics[width=2.25in]{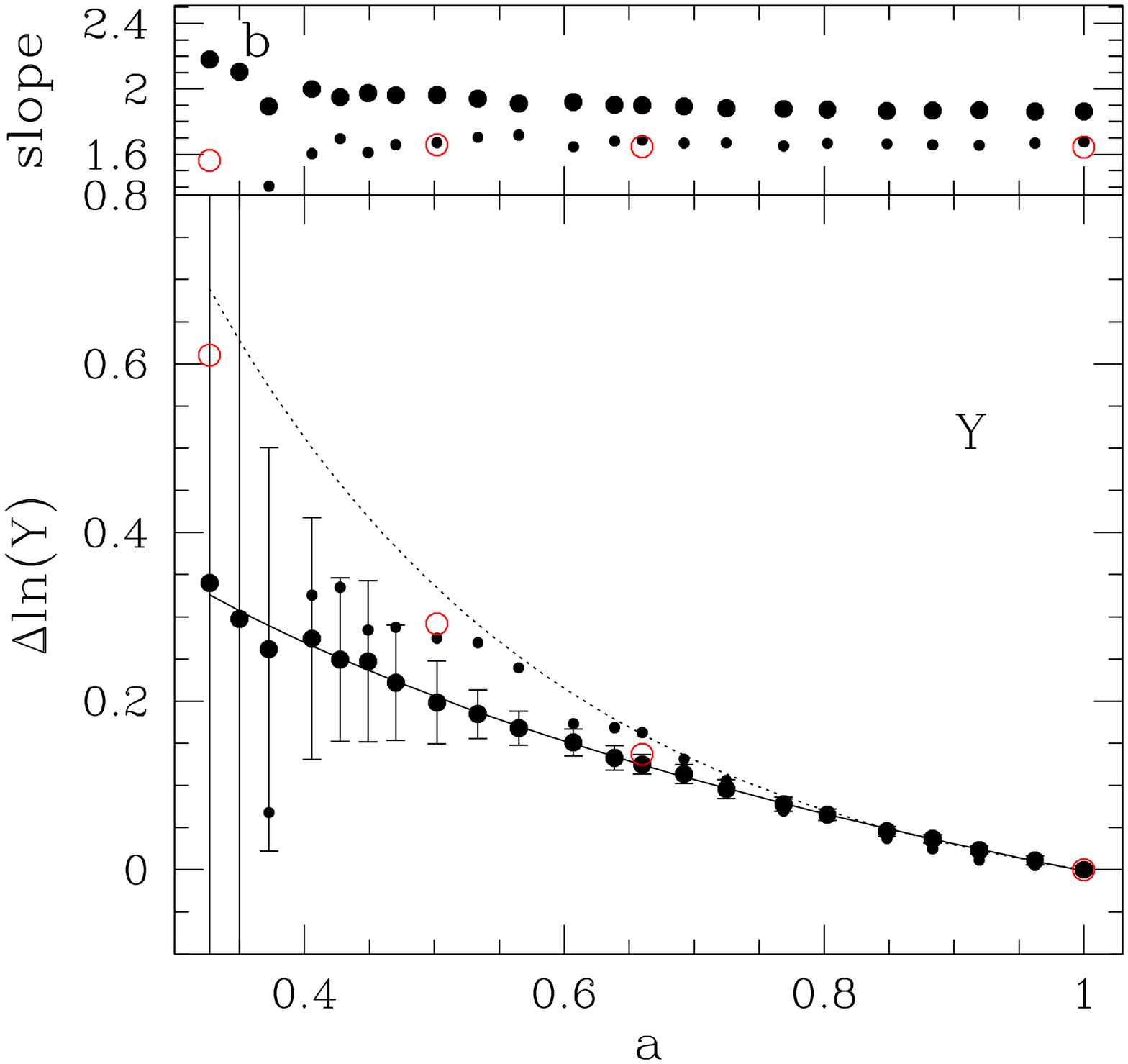} \\
\includegraphics[width=2.25in]{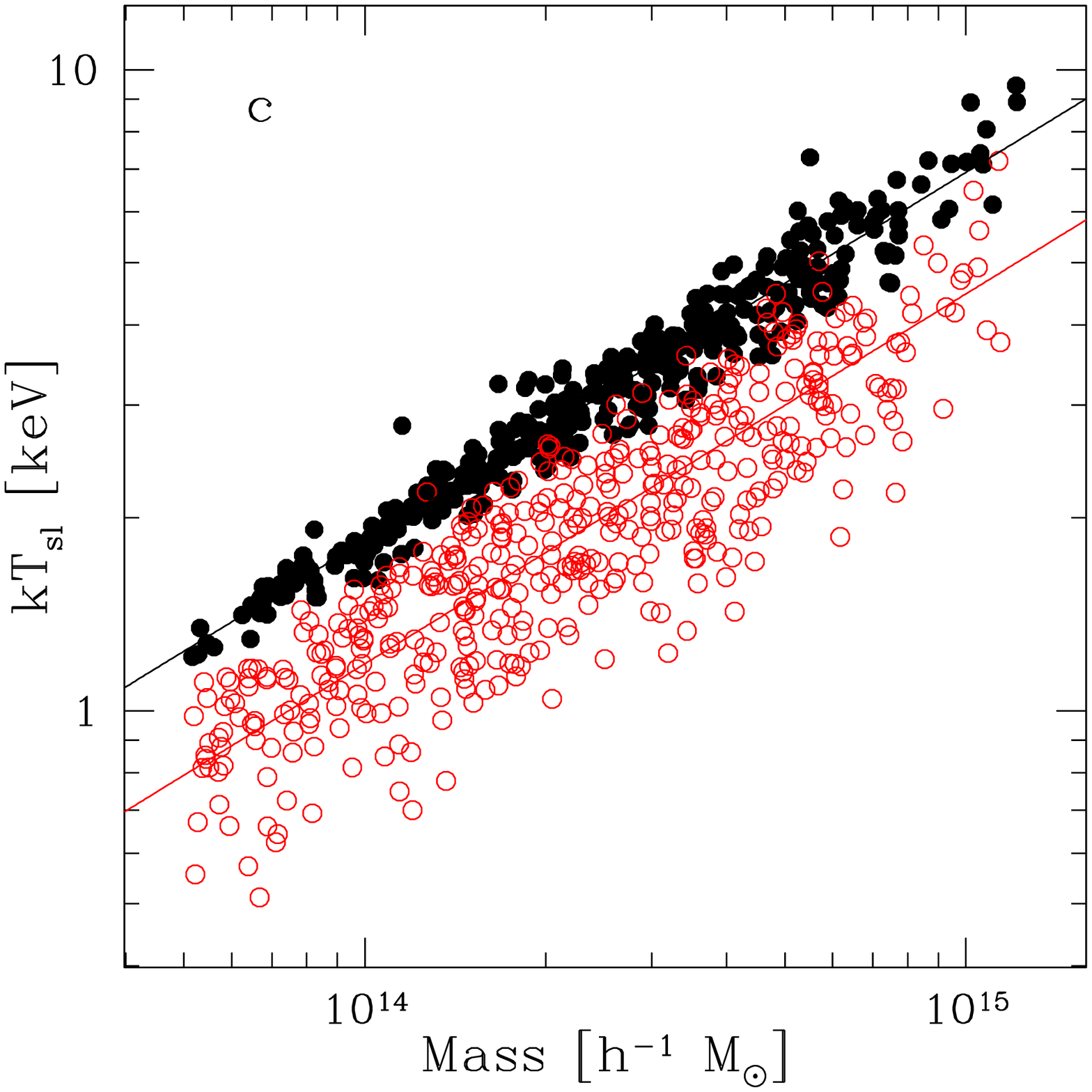} &
\includegraphics[width=2.25in]{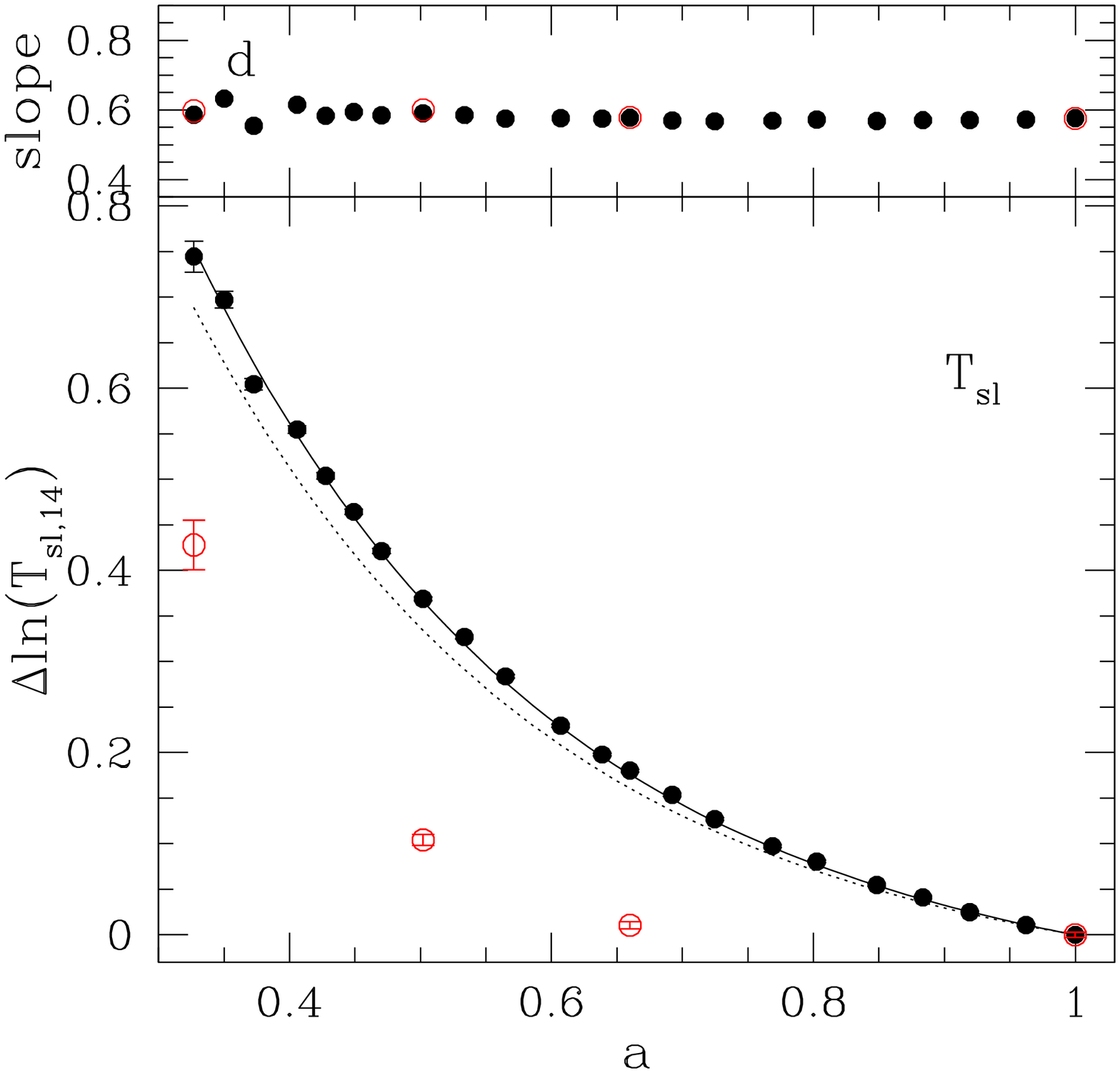} \\
\includegraphics[width=2.25in]{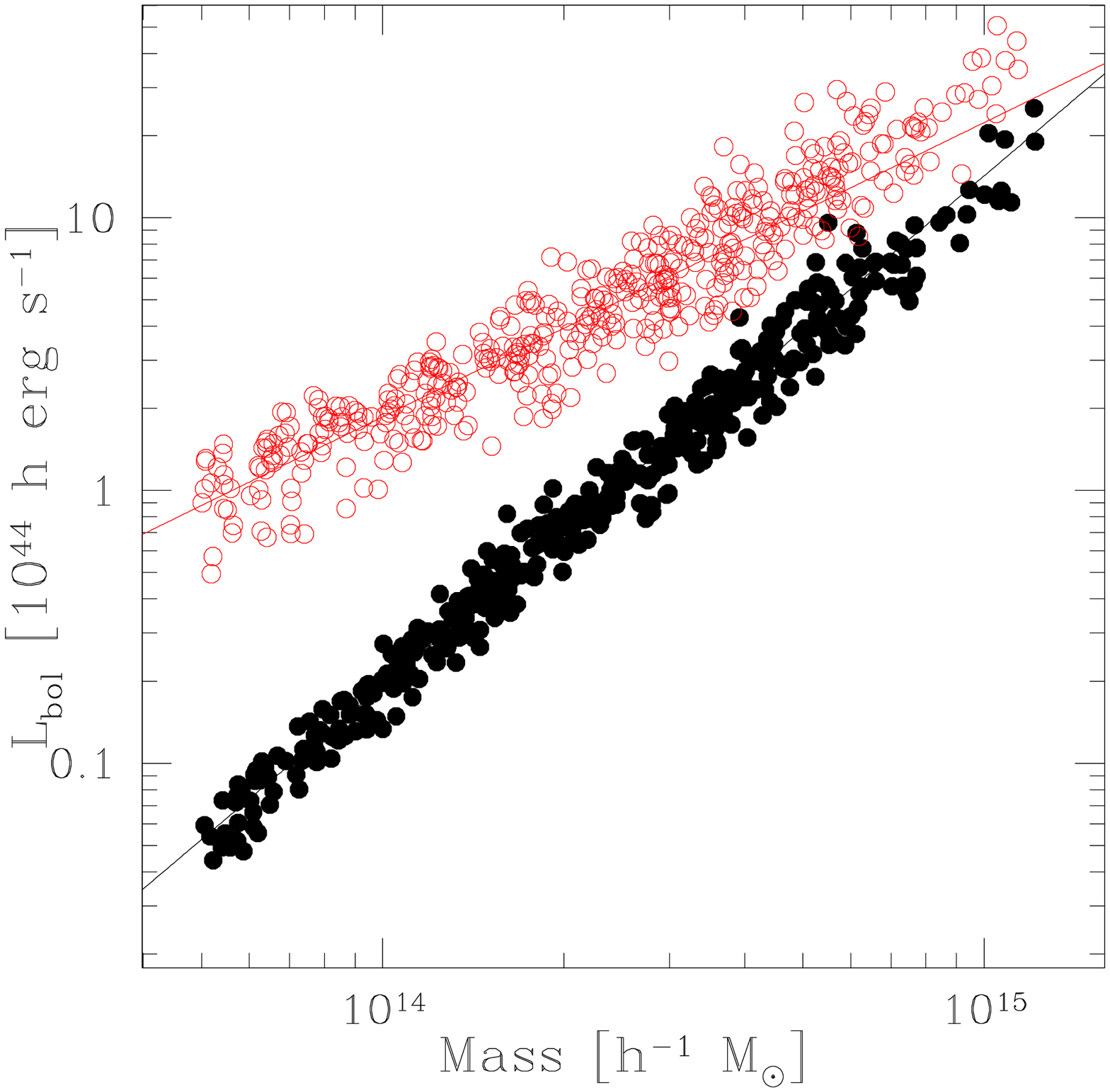} &
\includegraphics[width=2.25in]{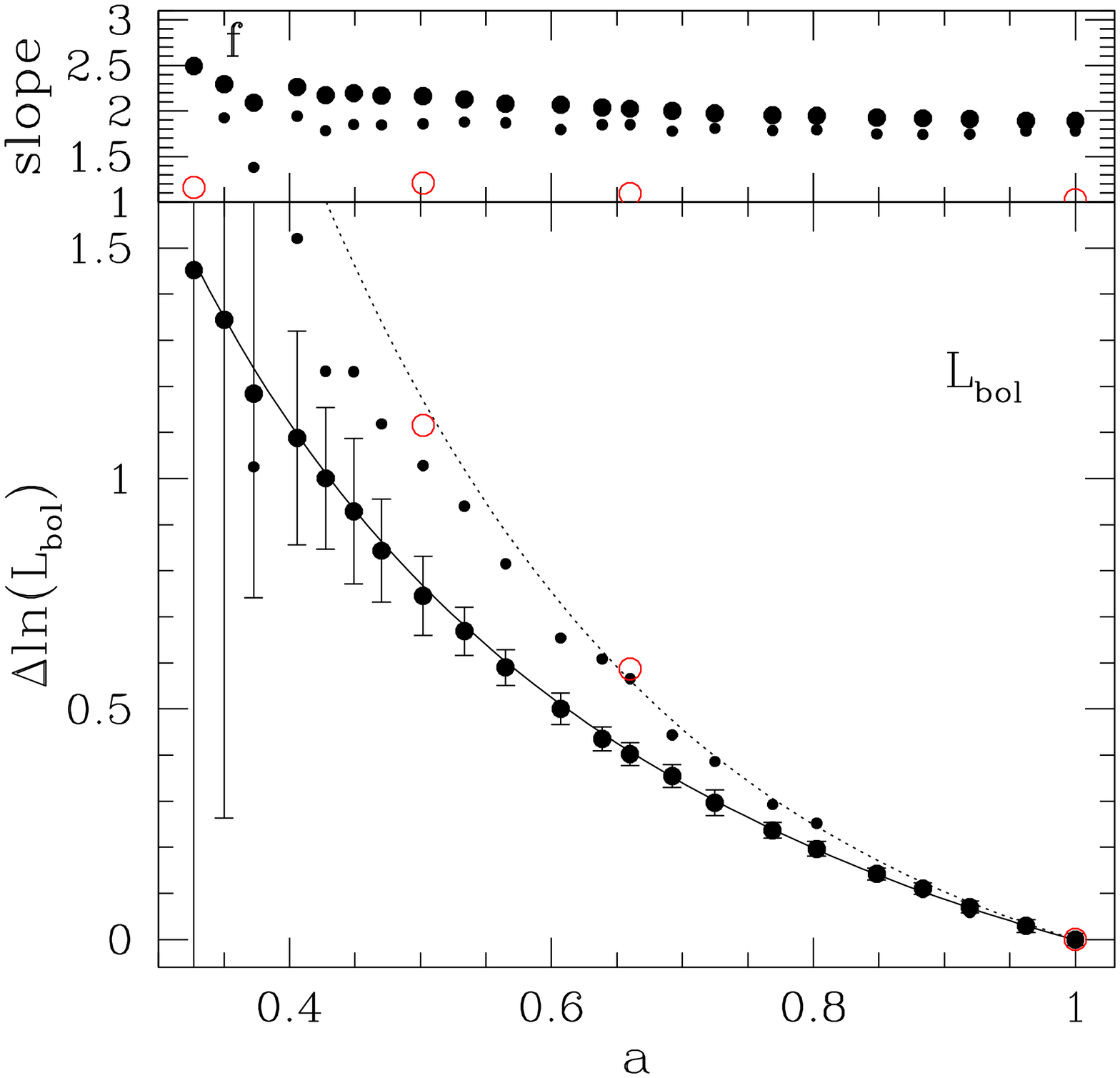} \\
\includegraphics[width=2.25in]{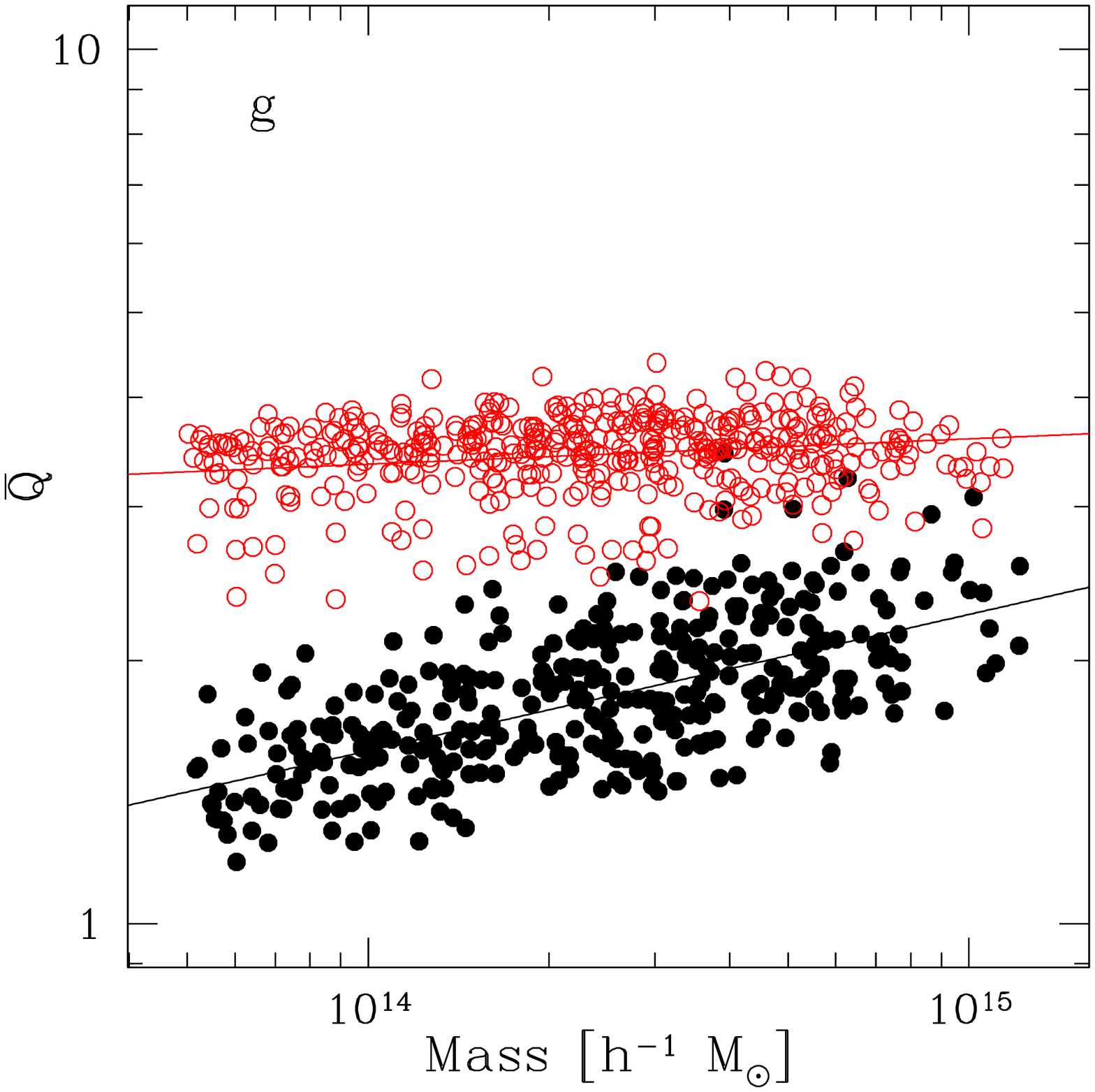} &
\includegraphics[width=2.25in]{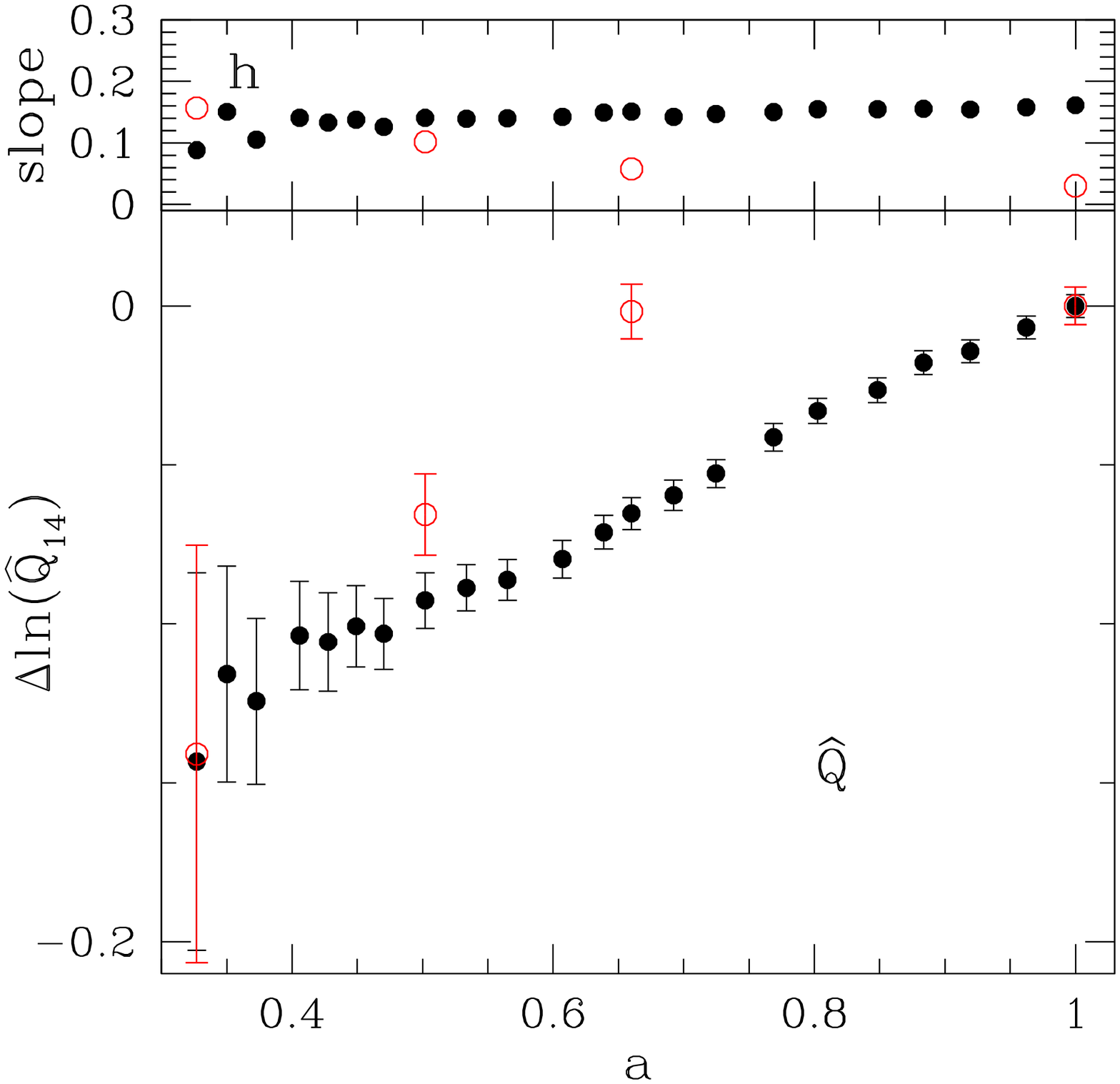}
\end{array}$
\end{center}
\caption{
Scaling relations at redshift zero (left) and redshift evolution of fit parameters (right) for $Y$, 
$\tsl$, $\elbol$, and $\hat{Q}$ (top to bottom).   Point and line styles are identical to Figure~\ref{fig:scalestructure}.   
\label{fig:scalexray}}
\end{figure*}

{\it Sunyaev-Zeldovich decrement.}  From Figure~\ref{fig:scalestructure}c,e, we know that preheated halos at fixed mass have, on average, lower ICM mass fractions and higher mass-weighted temperatures than their \GO counterparts.  Since the integrated thermal SZ decrement, $Y$, is a product of these two measures, we can anticipate some degree of cancelation coming from this  opposing behavior.  Figure \ref{fig:scalexray}a shows that this cancellation is quite close to exact above a mass scale of $2 \times 10^{14} \hinv\msun$.  

The slope in the \GO model is very close to the self-similar expectation of $5/3$, but curvature in $\ficm$ tilts the \PH relation from a local slope of $-1.8$ to $-1.6$ across the mass range $10^{14}-10^{15} \hinv\msol$.  As with the $\ficm-M$ relation, we fit $Y$ to a quadratic in $\ln M$ for the \PH case in order to account for the curvature and to get the best possible measure of the intrinsic scatter about the mean relation.  Parameters are given in Table \ref{tab:ficm}.

As discussed further in \S\ref{sec:covar}, the dispersion about the intrinsic, mean $Y-M$ relation is $12 \pm 1$ percent in both models.  The tight mass scaling is independent of cluster dynamical state, as noted by \citet{kravtsov:06},  making the SZ thermal signal ideal for cluster detection.  It must be remembered that this scatter is enlarged by line-of-sight projections, and that the magnitude of this effect will be model-dependent \cite{springel:01,shaw:06,hallman:07}.  

While the $Y-M$ normalization at high masses is not sensitive to cluster physics in our models, previously ART simulations with cooling, star formation and supernova feedback (CSF) display a normalization drop by $25\%$ relative to the \GO case  \citet{nagai:06}.  At a basic level, the different behavior between the two studies simply reflects the fact that the thermal pressure support, $-\nabla P/\rho$, is insensitive to a multiplicative shift in gas density normalization.  Our models force all baryons into the hot phase, while a CSF treatment allows some fraction of baryons --- 40\% in the case of  \citet{nagai:06} ---  to reside in galactic sinks of stars and cold gas.  Since observations indicate that the ratio of stellar to hot gas mass declines with increasing mass, from values near $0.5$ at $10^{14}\hinv\msol$ to $0.1$ at $10^{15} \hinv\msol$ \citep{giodini:09}, one would anticipate that the observed $Y-M$ relation to be steeper than self-similar.  A recent X-ray based estimate of the $Y-M$ relation from a {\sl Chandra} archival sample of groups and clusters finds a slope of $1.75$ at $r_{500}$ over the mass range $10^{13}-10^{15} \hinv\msol$ \citep{sun:09}. 

Figure ~\ref{fig:scalexray}b compares the redshift evolution of the $Y-M$ relation in the two simulations.  The slope in the \PH model steepens at higher redshift, and the normalization drifts below self-similar expectations by $\sim 10\%$ at $z =1$.  The evolution at high mass ($\sim 5 \times 10^{14} \hinv \msun$) in the \PH model is similar to that of the \GO model, in both slope and normalization, since redshift $z \sim 1$.  At our chosen normalization mass of $\sim 10^{14} \hinv \msun$, the evolution departs from self-similar at $z > 0.5$, 
driven by the evolution of the baryon fraction at that scale.  Like the baryon fraction, we fit the evolution of $Y$ to a quadratic in $\ln(a)$, with the best fit presented in Table \ref{tab:quadevol}.  We note that the \PH behavior is mildly in conflict with the CSF model evolution of \citet{nagai:06}, which displayed consistency with self-similar evolution at the $\sim 20\%$ level.  With only 11 halos, that study could not address statistical differences at the level we do here.   

{\it Spectroscopic-like temperature.}  We also consider the spectroscopic-like temperature, $\tsl$, an analytic prescription
 derived by \citet{mazzotta:04} to approximate observed X-ray spectral temperatures.  We find 
 %(Rasia \etal, in prep) 
that $\tsl$ agrees well with the X-ray temperatures derived from spectral fits to X-MAS2 mock observations \citep{rasia:05}.  As seen in Figure ~\ref{fig:scalexray}c,d, the slopes in the \PH and \GO treatments agree well, $\alpha = 0.57$, but at fixed mass the \PH halos are $\sim 40\%$ hotter than the \GO halos.  The distribution of the gas in the \GO models contains cool, low entropy cores of accreted sub-halos \citep{mathiesen:01}, and these cool, dense clumps pull down the $\tsl$ measure relative to the \PH simulation.  The cores in the latter case have been effectively erased by the preheating.  

The redshift evolution of $T_{sl,14}(a)$ in the \PH simulation is very close to self-similar ($\propto [E(a)]^{2/3}$), and tracks well the mass-weighted behavior shown above (see Table \ref{tab:evol}).  The evolution in the \GO simulation departs dramatically from self-similar behavior at low redshifts.

{\it Bolometric luminosity and Emission measure.}  The different behaviors of the baryon fraction and temperature in the \PH and \GO simulations drive X-ray luminosity differences, but the gas clumping, or emission measure, also plays a significant role.  Figure ~\ref{fig:scalexray}e,g show the $z=0$ mass scalings of the bolometric luminosity, $\elbol$, and dimensionless emission measure, $\hat{Q}$.   Because of the influence of line emission, the \GO scaling is less steep than the self-similar slope of $L \sim M^{4/3}$, and the emission measure is nearly constant with mass at a value $\sim 3.5$.  In the \PH case, the clumping factor is smaller by a factor of two and displays a significant trend with mass.   Combined with the $\ficm$ behavior, the result is a suppression of $\elbol$ in the \PH  case by a factor of 10 at $10^{14} \hinv\msol$, and a steepening of the slope in mass to $1.87$, from $1.08$ in the \GO case.  

Although the \PH halos have a higher temperature at fixed mass, this effect is dwarfed by the decreases in $\ficm$ and $\hat{Q}$ between the \PH and \GO treatments. The lower normalization of the $\hat{Q}-M$ relation reflects a shallower gas density profile in the \PH case.  In turn, the lower central gas densities contribute to lowering the overall mass profile, driving the shift in concentrations to lower values for the \PH halos discussed above.   As in the $Y-M$ relation, the curvature 
in the $\ficm-M$ relation in the \PH model drives curvature in the $\elbol-M$ relation.  \cite{hartley:08} see this curvature in the $L-T$ relation in the \PH 
model, as well as in a large, local observed sample of galaxy clusters.  We fit the $\elbol-M$ relation in the \PH model to a quadratic in $\ln M$, presenting the best 
fit in Table \ref{tab:ficm}.

% In Section \ref{sec:scatter} we consider how the scatter in the $\elbol-M$ relation depends on the scatter in $\ficm$, $T$, and $Q$ at fixed mass.

The evolution of the normalization in \PH is not a perfect power of $E(a)$, due to 
the complicated evolution of $\ficm$ with redshift.   We fit the evolution of $L_{14}(a)$ to a quadratic in $\ln(a)$ in the \PH simulation, and note that it 
is weaker than the \GO evolution.  However, as shown in Figure \ref{fig:scalexray}f, this evolution is a function of mass.  The larger solid points, for halos 
at $10^{14} \hinv \msun$, show a weaker evolution than the evolution of the halos at $5 \times 10^{14} \hinv \msun$, shown by the smaller solid points.  
The latter halos evolve similarly to the \GO halos, illustrating that the most massive halos in the \PH model are similar in structure and history to the 
\GO population.  The evolution of 
the $\elbol-M$ relation in the \PH simulation is driven mainly by the redshift evolution of the hot gas fraction, with $\hat{Q}$ contributing $10\%$ of the decrease at $z=1$.

%%%%%%%%%%%%%%%%%%%%%%%%%%%%%%%%%%%%%%%%%%%%%%%%%%%%%%%%%%%%%%%%%%%%%%%
\subsection{Comparison to Observations}\label{sec:meanobs}

As shown by \cite{hartley:08}, the \PH simulation offers a good match to the core-excised $L-T$ relation  of local clusters.   Here, we briefly explore the level of agreement between the models and observations in this and other scaling relations.  While not wishing to oversell the simple physical treatment of the preheated model, which is undoubtedly wrong in detail, we show below that it reproduces several scalings with quite high fidelity.  Preheating appears to be a useful effective model.  It is important to remember that precise comparison of observations and simulation expectations requires careful modeling of survey selection and projection effects, and we do not treat these effects here.  

\begin{figure}
\plotone{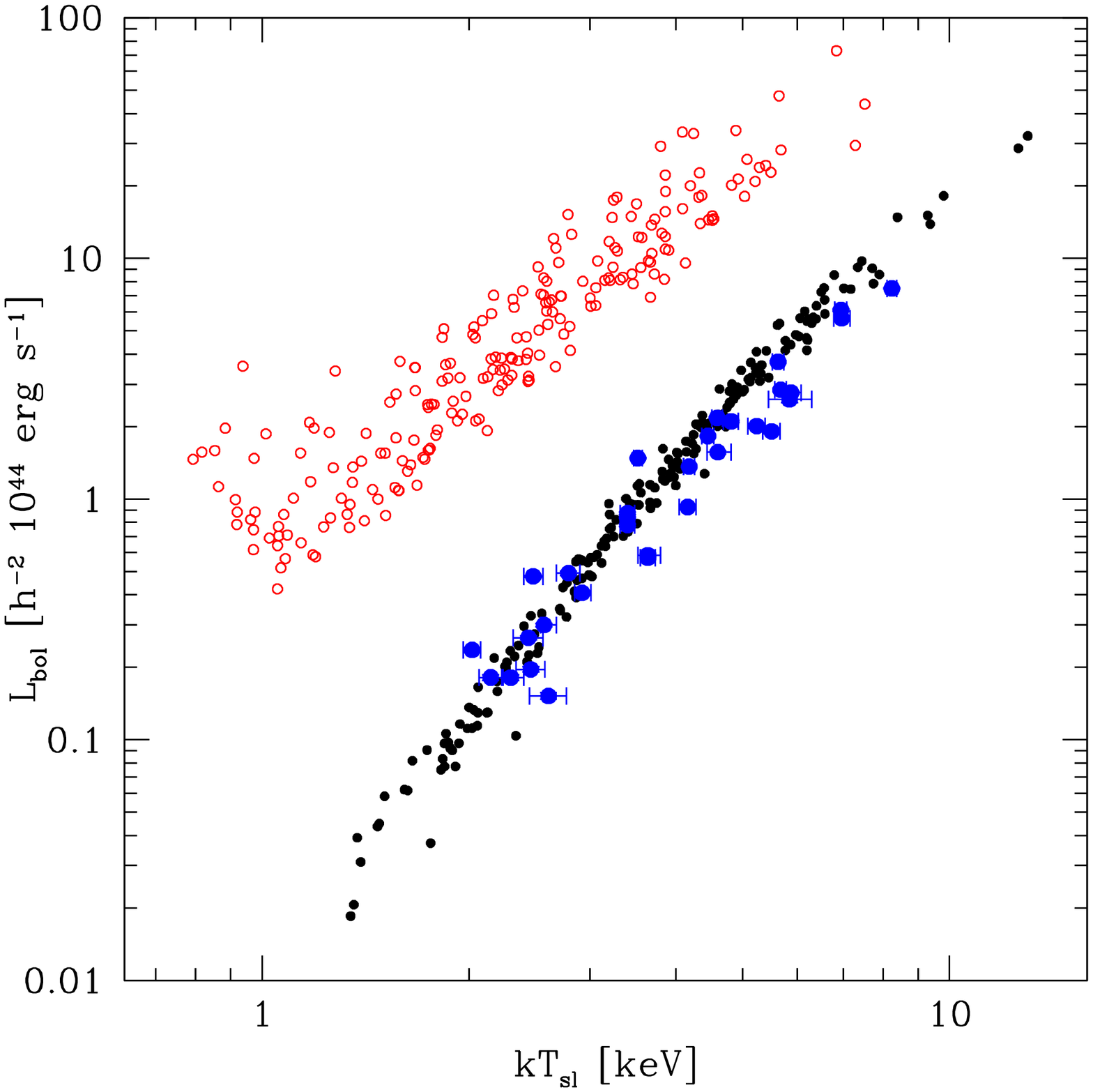}
\caption{The redshift zero $\elbol-\tsl$ relation for \GO (open, red points), \PH (filled, black  points), measured within 
$\Delta_c = 500$, and the core-excised $\elbol-T$ relation from the REXCESS survey (large, blue points) 
\citep{pratt:08}.
\label{fig:ltrexcess}}
\end{figure}

Figure \ref{fig:ltrexcess} compares the $\elbol-\tsl$ relations with the core-excised $\elbol-T_X$ measurements of the local, representative REXCESS survey \citep{pratt:08}.   The slope of the observed relation is somewhat shallower, and its scatter somewhat larger, than the \PH model predictions, but these differences are at the level of a few tens of percent in luminosity, or less than ten percent in temperature.  
The small scatter in the core-excised $\elbol-T$ relation for REXCESS clusters (and \PH halos) indicates that local 
galaxy clusters are well-behaved outside of the core \citep{neumannArnaud:01}.   
% Simulations by \citet{dolag:05} find similar outer structure for halos simulated with different physical treatments for the ICM.   

\begin{figure}
\plotone{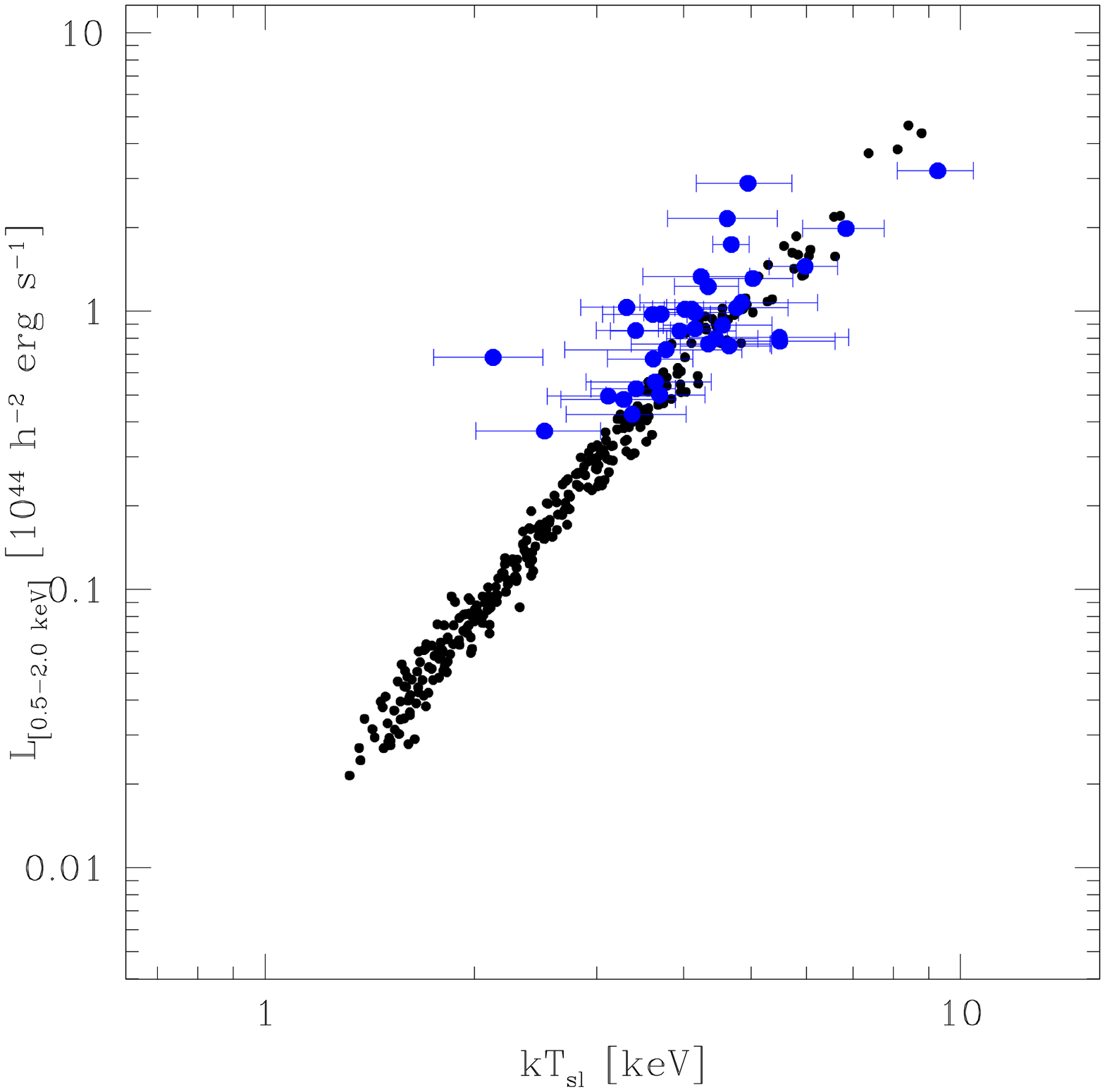}
\caption{The redshift $z = 0.5$ $L_{[0.5-2.0]}-\tsl$ relation for \PH (filled, black  points), measured within 
$\Delta_c = 500$, and the CCCP $L_{[0.2-2.0]}-T$ relation from \citet{vikhlinin:08} (large, blue points), scaled to redshift $z = 0.5$.
\label{fig:ltcccp}}
\end{figure}

At higher redshift, we consider the CCCP clusters, a subset of the 400 square degree survey which 
has been followed-up with Chandra \citep{vikhlinin:08}.  These clusters range from approximately 
$0.3 < z < 0.8$, so at fixed mass we scale, in a self-similar manner, the observed luminosities and temperatures to $z = 0.5$ for comparison with the model $\elbol-\tsl$ relations. The latter are measured within $\Delta_c = 500$ to be consistent with the treatment of \cite{vikhlinin:08}.  The comparison is presented In Figure~\ref{fig:ltcccp}.   The agreement is good, but the measurement errors are larger than the REXCESS sample.  The scatter in the CCCP sample is larger, and this may be due in part to the fact that cores have not been excised in the luminosity measurements of this sample.  

\begin{figure}
\plotone{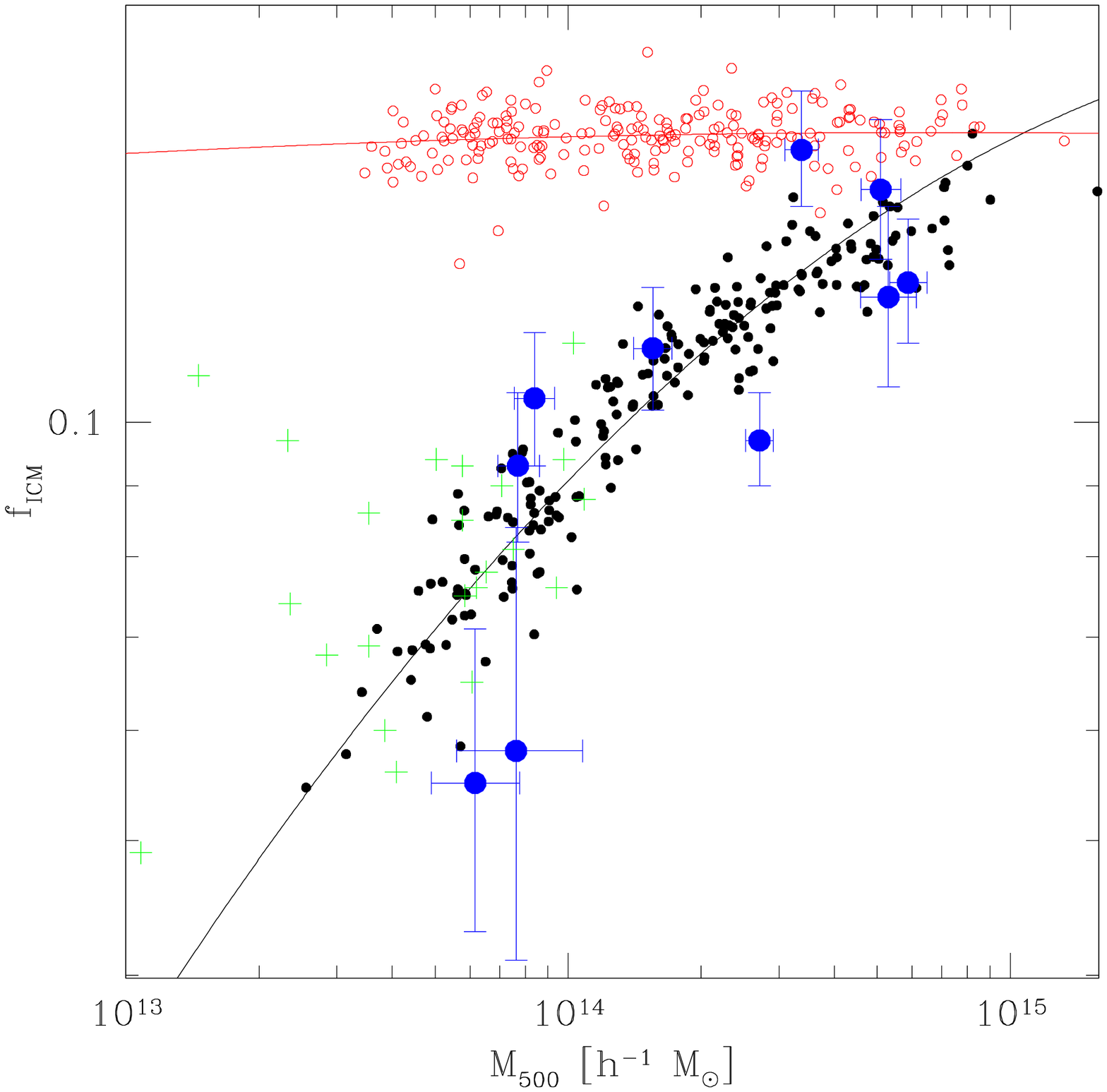}
\caption{The redshift zero $\ficm-M$ relation for \GO (open, red points), and \PH (filled, black  points), measured within 
$\Delta_c = 500$. The observations plotted are the data from \citet{arnaud:07} (large, blue points) and from 
\cite{sun:09} (green crosses).
\label{fig:ficmarnaud}}
\end{figure}

Figure \ref{fig:ficmarnaud} compares the mass scaling of ICM mass fractions at redshift zero in the models to XMM measurements for local clusters \citep{arnaud:07, sun:09}.   Values are measured within $\Delta_c = 500$.   To estimate observed cluster masses, hydrostatic estimates that include a radial temperature gradient are employed.  We note that Arnaud data agree with other observational determinations at high mass \citep{vikhlinin:06, giodini:09}, while the  \cite{sun:09} data extend to lower mass, $\sim 10^{13} \hinv \msun$.  The \PH model matches the observed  $\ficm-M$ relation well, with hot gas fractions a factor two less than the cosmic ratio, $\Omega_b/\Omega_m$ at $10^{14}$ and a strongly increasing trend toward higher masses.  Larger observed samples are needed to test the curvature and degree of scatter.  

\begin{figure}
\plotone{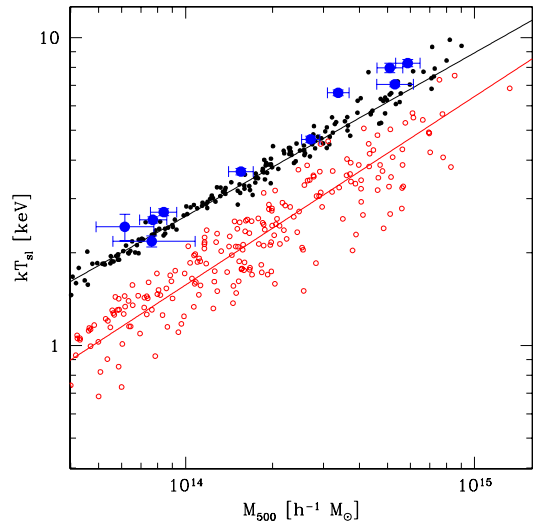}
\caption{The redshift zero $\tsl-M$ relation for \GO (open, red points), \PH (filled, black points), measured within 
$\Delta_c = 500$, and the data from \citet{arnaud:07} (large, blue points).
\label{fig:tslarnaud}}
\end{figure}

In Figure \ref{fig:tslarnaud}, we compare the $\tsl-M$ relation of the models to that from \cite{arnaud:07}.   The \GO model, with its cool sub-halo cores, is strongly offset from the data.  The \PH model relation is much closer, with a similar slope and scatter.  There is, however,  a consistent offset of $\sim 15\%$ in mass toward lower values in the observed sample.   The magnitude of this offset is consistent with the level of expected bias from hydrostatic mass estimates, which simulations show tend to underestimate true masses by approximately $20\%$ \citep{rasia:06, nagai:07}.

Overall, the bulk X-ray properties of the \PH simulation match the local scaling relations quite well.  This is particularly 
true after excising the core from the observations, and after considering the mass bias introduced by 
hydrostatic mass estimates.  Although the physics of the \PH model is simple, it appears to provide useful representation for the behavior of the bulk of the hot ICM lying outside the cool core regions.

%%%%%%%%%%%%%%%%%%%%%%%%%%%%%%%%%%%%%%%%%%%%%%%%%%%%%%%%%%%%%%%%%%%%%%%%%
%%%%%%%%%%%%%%%%%%%%%%%%%%%%%%%%%%%%%%%%%%%%%%%%%%%%%%%%%%%%%%%%%%%%%%%%%
\section{Covariance of Bulk Properties} \label{sec:covar}

In this section, we explore the second moment of the halo scaling relations.   Understanding the variance at 
fixed mass is necessary for calculating the mass selection properties of signal-limited samples, and survey counts typically constrain only a linear combination of signal normalization and variance \citep{stanek:06}.  In addition, the signal covariance determines the precise structure of scaling relations in signal-selected samples.  A worked example of the $L-T$ relation expected for X-ray flux-limited samples is given by \cite{nord:08}.  

After defining terms, we begin by presenting the covariance of signals at fixed mass at the present epoch, then demonstrate that redshift evolution in most elements is weak.  We close with analysis of the mass selection properties of signal pairs.  

\subsection{Signal Covariance Matrix}\label{sec:covardefn}

For a set of halos of mass $e^\mu$ at expansion epoch $a$, we define a symmetric covariance matrix, with elements 
\begin{equation} 
\Psi_{ij}  \ \equiv  \ \langle  (s_i-\bar{s}_i(\mu,a)) (s_j-\bar{s}_j(\mu,a)) \rangle
% \Psi_{xy} = \frac{1}{N} \sum_i^N (x_i-\bar{x}(\mu,a))(y_i-\bar{y}(\mu,a)),
\label{eqn:psidefn}
\end{equation}
where the mean values are determined by equation~(\ref{eqn:lsqfit}) and the brackets represent an ensemble average.   
The $j^{th}$  diagonal element of the covariance matrix is the $j^{th}$ signal variance, $\sigma_j^2 \equiv \Psi_{jj}$.   We present these measures along with the correlation matrix, $C_{xy} =  \Psi_{xy} / (\sigma_x \sigma_y)$, which expresses the covariance in normalized terms.  

\begin{figure*}
\plotone{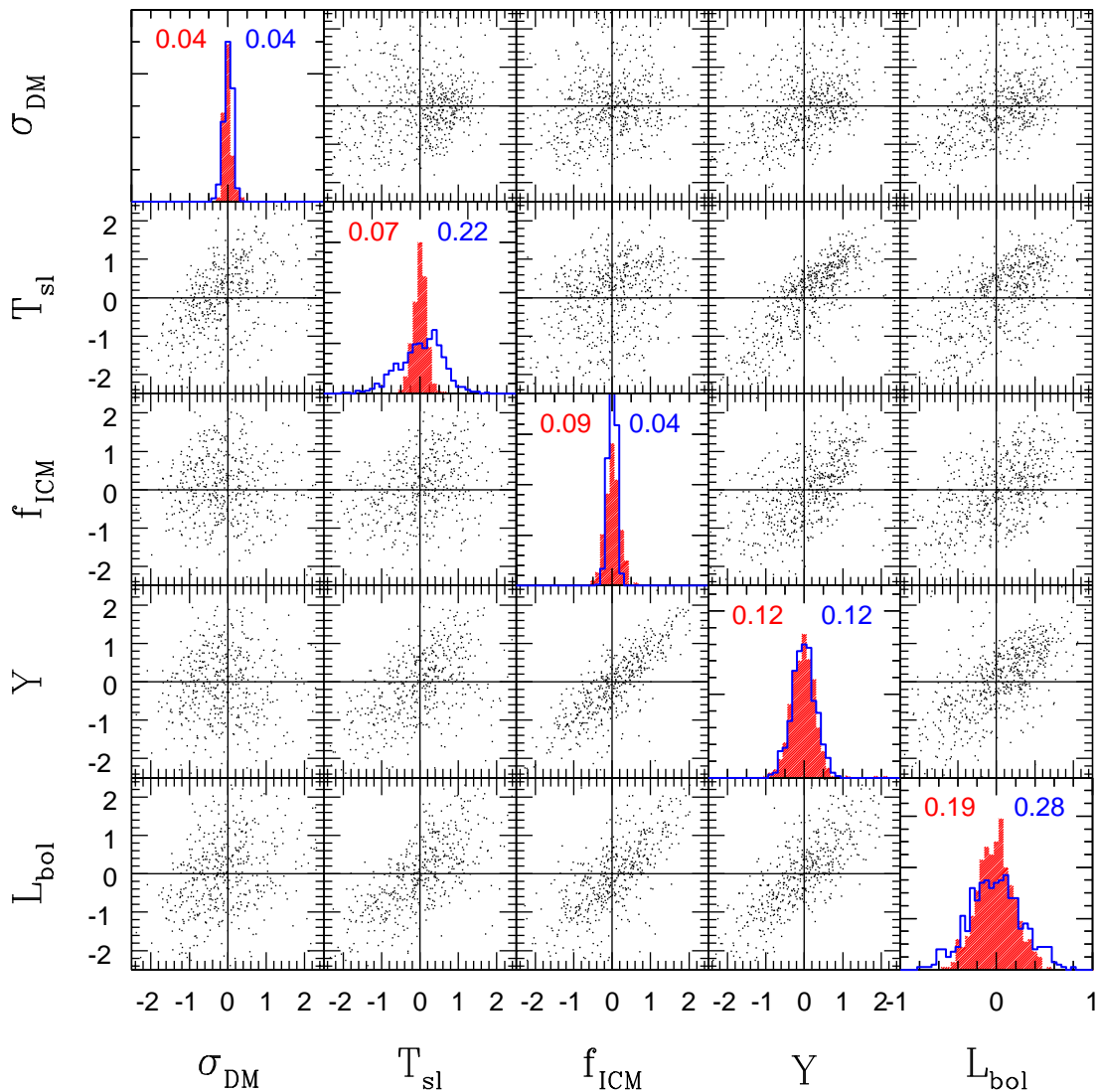}
\caption{Graphical representation of the data comprising the covariance matrix for the two simulations at  
$z = 0$.  Diagonal panels plot the distribution of deviations from mean mass scaling behavior, $(s_i-\bar{s}_i)$, where $s_i$ is the natural log of the $i^{\rm th}$ signal.   Shaded histograms show \PH results and solid lines show \GO data. 
Each off-diagonal panel plots the normalized deviations, $(s_i-\bar{s}_i) / \sigma_i$, for an $(i,j)$ pair of properties.  The lower triangle shows \PH and the upper triangle shows \GO behavior.  
\label{fig:covarz0}}
\end{figure*}

Figure \ref{fig:covarz0} presents a graphical representation of the data comprising the covariance matrix for a subset of signals at redshift zero.   The diagonal panels show the distribution of signal deviations (in the natural log of the measured signal) about the mean for the \PH (shaded) and \GO (lines) treatments.   Panels off the diagonal plot the normalized deviations 
($(s_j-\bar{s}_j) / \sigma_j \ {\rm vs.} \ (s_i-\bar{s}_i) / \sigma_i$) for signal pairs, with the lower and upper triangles showing \PH and \GO cases, respectively.  The orientation and spread of halos in each panel determines the correlation coefficient.  For instance, the tight ellipse formed by the population in the $Y-\ficm$ panel indicates a high correlation coefficient between this pair of signals.  

The $z=0$ values of the correlation coefficients, along with uncertainties from bootstrap resampling, are presented in Table \ref{tab:r0}.  As in the mean fit parameters, typical uncertainties are on the order of $\sim 1\%$.  Before exploring the off-diagonal terms, we first examine the variance and distribution function shape of individual signal deviations.

%%%%%%%%%%%%%%%%%%%%%%%%%%%%%%%%%%%%%%%%%%%%%%%%%%%%%%%%%
\subsection{Signal Variance at Fixed Mass}\label{sec:scatter}

The assumption of log-normal variance is a common element of the likelihood analysis used in cluster cosmology studies (see references in \S1).  Both the absolute variance and the full distribution shape affect the expected signal counts derived from mass function convolution.   

In Table \ref{tab:quint}, we list root mean square deviations for the full set of signals at $z=0$.  To test the log-normal expectation, we list the normalized deviates at which each signal's ranked distribution reaches fixed percentile values, taken to be $\pm 2\sigma$, $\pm 1\sigma$ and median/mean of a Gaussian distribution.   The difference between the listed values and their integer counterparts is a measure of the degree of local deformation from Gaussian in the frequency distribution.

\begin{deluxetable*}{l|c|ccccc}
\tablecaption{\label{tab:quint}Scatter and Distribution of Deviates at Redshift Zero\tablenotemark{a}}
\tablehead{
\colhead{Signal} & \colhead{Scatter} & \colhead{$2.275\%$} & \colhead{$15.865\%$} & \colhead{$50\%$} & 
\colhead{$84.135\%$} & \colhead{$97.725\%$}
}
\startdata
Gaussian &- &  -2.00 & -1.00 & 0.00 & 1.00 & 2.00 \\ \hline
\PH $\sigmadm$ &$0.042 \pm 0.001$ & -1.66 & -0.90 & -0.11 & 0.85 & 2.43 \\
\PH $\ficm$ &$0.086 \pm 0.001$ & -2.14 & -0.93 & 0.03 & 0.93 & 1.93 \\
\PH $T_m$ &$0.058 \pm 0.002$ &  -1.90 & -0.84 & -0.04 & 0.75 & 2.11\\
\PH $\tsl$ & $0.069 \pm 0.001$ & -2.17 & -0.93 & 0.04 & 0.90 & 1.88 \\
\PH $Y$ & $0.125 \pm 0.002$ & -2.04 & -0.97 & 0.04 & 0.90 & 1.95 \\
\PH $\elbol$ & $0.193 \pm 0.002$ & -1.98 & -0.98 & 0.004 & 1.00 & 1.94 \\
\PH $\hat{Q}$ & $0.116 \pm 0.001$ & -1.89 & -0.99 & -0.04 & 0.97 & 2.19\\
\PH $c$ & $0.300 \pm 0.008$ & -1.27 & 0.43 & 0.13 & 0.89 & 1.31 \\ \hline
\GO $\sigmadm$ & $0.042 \pm 0.001 $ & -1.65 & -0.88 & -0.12 & 0.77 & 2.54\\
\GO $\ficm$ & $0.036 \pm 0.001$ & -2.12 & -0.98 & 0.03 & 0.99 & 1.90 \\
\GO $T_m$ & $0.102 \pm 0.001$ & -2.23 & -1.00 & 0.12 & 0.93 & 1.69 \\
\GO $\tsl$ & $0.219 \pm 0.002$ & -2.32 & -1.06 & 0.19 & 0.96 & 1.49 \\
\GO $Y$ & $0.123 \pm 0.001$ & -2.17 & -1.00 & 0.10 & 0.95 & 1.75 \\
\GO $\elbol$ & $0.282 \pm 0.003$ & -2.19 & -1.03 & 0.10 & 0.97 & 1.70 \\
\GO $\hat{Q}$ & $0.109 \pm 0.001$ & -2.59 & -0.91 & 0.19 & 0.85 & 1.55 \\
\GO $c$ & $0.280 \pm 0.008$ & -2.42 & -0.97 & 0.13 & 0.86 & 1.28 
\enddata
\tablenotetext{a}{The distribution of deviates shows locations, in terms of  normalized deviates, $\delta/\sigma$, at which the ranked distributions reaches the listed percentiles.   Values for a Gaussian distribution are listed in the first row.
 for each signal.  Error estimates on the scatter come from bootstrap resampling. }
\end{deluxetable*}

While deviations from Gaussianity are apparent in essentially all measures, the  typical percentile shifts in the \PH model are only a few percent.   Exceptions are a significant positive skew in the dark matter velocity dispersion, with median location of $-0.12\sigma$ and shifts in the $\pm 2\sigma$ Gaussian tails to $-1.7\sigma$ and  $2.5\sigma$, respectively.   Most distributions are slightly leptokurtic, especially the mass weighted temperature.  Worth noting is the fact that the shapes of two important cluster selection observables, $Y$ and $\elbol$, do not deviate by more than $0.1$ from the Gaussian expectations.  

Under the \GO treatment, the shape of the dark matter velocity dispersion is the same as in the \PH case, showing positive skew at the $10\%$ level.   However, the shapes of the hot gas properties of \GO halos generally differ, to a slight degree, from the \PH shapes; all measures tend to be slightly skew negative.  The differences can be subtle, as close inspection of the $Y$ histograms in Figure~\ref{fig:covarz0} confirms.   

Comparing the \GO and \PH distributions, we conclude that a log-normal approximation is a fairly accurate description, but calculations demanding better than $\sim 10\%$ precision in shape will require an expanded treatment, either via direct Monte Carlo from simulations, or from analytic extensions, such as an Edgeworth series in Hermite polynomials \citep{kofman:93, shaw:09}.   We leave detailed analysis of the physical mechanisms driving these distribution shapes to future investigations.  In the case of $\sigmadm$, we have preliminary evidence that mergers drive the positive skew tail of the deviations; nearly all halos with a deviation of $> 3\sigma$ have undergone a merger since redshift $z = 0.2$.   

The amplitudes of the scatter in the set of signals range from a low of $0.036$ for $\ficm$ to a high of $0.28$ for $\elbol$, both under the \GO treatment.  For the \PH case, $\elbol$ is highest, at $0.19$ while $\sigmadm$ is lowest at $0.042$.  The latter value matches the \GO case, and both agree with the scatter derived from the ensemble value presented by \cite{evrard:08}.  While the dark matter virial scaling is robust to simple physical treatments for the baryons, \citep{lau:09} find that strongly dissipative baryon physics depresses the slope of the $\sigmadm-M$ relation by introducing mass- and redshift-dependent increases in halo velocity dispersion.      
 
By raising the halo sound speed, which drives the shock radius to larger values and lowers the Mach number of infalling material \citep{voit:02},  preheating leads to more thermally regular halo gas, with smaller scatter in mass-weighted temperature, $T_m$, compared to the \GO simulation.  Naively, in a virialized cluster dominated by only gravitational effects, one expects $\sigmadm^2 \sim T_m$, and $2 \sigma_{\sigmadm} \simeq \sigma_{T_m}$.  The values for the \GO simulation are close to this, with $\sigma_{T_m} \sim 
2.5 \sigma_{\sigmadm}$.  However, the scatter in $T_m$ is much lower in the \PH simulation, only $1.4 \sigma_{\sigmadm}$, or $5.8$\%.  

The scatter in $\ficm$ shows the reverse behavior.  In the \GO simulation, the scatter is very small, $3.6\%$, consistent with other gravity-only SPH simulations \citep{ettori:06, crain:07}.  When only gravity drives gas thermodynamics, the gas distribution traces the dark matter 
distribution very well.   The \PH model scatter of $8.6\%$ is more than twice that of the \GO case.  
However, the scatter in this model is mass-dependent; splitting the sample at $3 \times 10^{14} \hinv \msun$, the low-mass end has 
$\sim 13\%$ scatter while it is only $\sim 7\%$ on the high-mass end.   

The opposing statistical shifts in $\ficm$ and $T_m$ effectively cancel when combined to form the thermal SZ signal.  
Not only does the mean $Y-M$ relation in Figure~\ref{fig:scalexray} agree well between the two simulations, but both 
simulations have a similarly low scatter, $\sim 12\%$.   The small intrinsic scatter is consistent with results from previous simulations \citep{evrard:90b, dasilva:00, kay:04, Motl:05, nagai:06, kravtsov:06, hallman:07}.  As discussed below, the combination of low scatter in $Y$ and the steep slope of the $Y-M$ relation make the thermal SZ effect, or its X-ray equivalent, an excellent mass proxy for cluster surveys.  Furthermore, as discussed above, the distribution of deviations in $\ln(Y)$ at fixed mass is very close to a Gaussian distribution.   

In both models, the scatter in $\tsl$ at fixed mass is higher than the scatter in $T_m$ at fixed mass. The difference 
is slight in $\PH$, but over a factor of two in \GO, a reflection of the larger amount of cool substructure in the latter treatment.  
The fact that $\tsl$ is sensitive to the gas physics treatment means that opportunities exist for constraining gas physics based on high-quality X-ray spectroscopy of large samples.  Such a study could be provided by the proposed WFXT satellite\footnote{http://wfxt.pha.jhu.edu/}.

After NFW concentration, the X-ray luminosity has the highest level of scatter at fixed mass in both the \PH ($\sigma_{\ln L}=0.19$) and \GO ($\sigma_{\ln L}=0.28$) simulations.    These values are consistent with the  core extracted value, $\sigma_{\ln L} = 0.27 \pm 0.06$, observed for REXCESS \citep{pratt:08}. 

%%%%%%%%%%%%%%%%%%%%%%%%%%%%%%%%%%%%%%%%%%%%%%%%%%%%%%%%%%%%%%%%%%%%%%%%%%
\subsection{Off-Diagonal Elements of the Correlation Matrix}\label{sec:offdiag}

\begin{deluxetable*}{lcccccccc}
\tablecaption{Correlation Coefficients at Redshift Zero \label{tab:r0}\tablenotemark{a}}
\tablehead{
\colhead{Signal} & \colhead{$\sigmadm$} & \colhead{$T_m$} & $\tsl$ 
& $\ficm$ & $Y$ & $\elbol$ & $\hat{Q}$ & $c$}
\startdata
$\sigmadm$ & $ -$ & $0.55$ & 0.81 & 0.28 & 0.54 & 0.51 & 0.17 & 0.19 \\
$T_m$ & 0.35 & $ - $ & 0.85 & 0.48 & 0.97 & 0.67 & 0.38 & 0.49 \\
$\tsl$ & 0.86 & 0.5 & $ - $ & 0.42 & 0.83 & 0.67 & 0.47 & 0.64 \\
$\ficm$ & -0.10 & 0.42 & 0.37 & $ - $ & 0.69 & 0.60 & 0.32 & 0.37 \\
$Y$ & 0.079 & 0.74 & 0.62 & 0.88 & $ - $ & 0.73 & 0.40 & 0.51 \\
$\elbol$ & 0.26 & 0.50 & 0.73 & 0.76 & 0.78 & $ - $ & 0.65 & 0.70 \\
$\hat{Q}$ & 0.32 & 0.029 & 0.56 & 0.15 & 0.12 & 0.59 & $ - $ & 0.71 \\
$c$ & 0.15 & 0.053 & 0.39 & 0.29 & 0.26 & 0.51 & 0.64 & $ - $ 
\enddata
\tablenotetext{a}{The redshift zero correlation coefficients, with the results from the \PH 
simulation in the lower triangle and the results from the \GO simulation in the upper, as in Figure \ref{fig:covarz0}.   
Uncertainties from bootstrapping resampling are on the order of $0.01$ and are not shown.}
\end{deluxetable*}

There is much information about the physical processes driving the ICM encoded in the correlation matrix of Table \ref{tab:r0}.  
For cosmological studies, it would be useful to identify pairs of cluster properties whose 
correlation coefficient is insensitive to gas physics modeling.   
High values of the correlation coefficient may point to pairs of halo properties that evolve on similar time scales during merger events.  Below in \S{sec:imp}, we show that high signal correlations can improve halo mass selection by joint signals considerably.  

Overall, the signal pair correlations tend to be higher in \GO than in \PH, but detailed differences warrant closer inspection.  Considering the virial scaling as an anchor, we begin with consideration of correlations with $\sigmadm$.  We then examine SZ and X-ray observables, and point out extreme values for both cases.   

The covariance between $\sigmadm$ and $T_m$ at fixed mass is an indicator of halo virialization.  At redshift zero, this  
correlation is much higher, $C = 0.56$ in \GO than in \PH, where it is $0.35$.  The halos in the \GO simulation are governed by gravitational effects only, so $T_m$ and $\sigmadm$ excursions track each other closely.  In \PH, however, the preheating raises the sound speed in the halos, making the thermalization due to mergers less pronounced, thereby diminishing (though not eliminating)  the coupling of $T_m$ and $\sigmadm$ deviations.

The ICM mass fraction behavior in the two simulations is quite different, as noted above, and this difference is apparent in the covariance of $\sigmadm$ and $\ficm$.   
% In the \PH simulation, the baryon fraction of halos is still increasing at 
% redshift zero, especially at low mass.  In the \GO simulation, however, there is no 
% trend with mass in $\ficm$, and only the slightest evolution since $z = 1$.  These 
% very different trends in evolution  strongly affect
 % the covariance between $\sigmadm$ and $\ficm$ at fixed 
% mass.  
In the \PH simulation, the two properties are anti-correlated, $C = -0.10$ 
at redshift zero.  In fact, this is the only negatively correlated pair of signal deviations exhibited by the models.  It is likely that this negative correlation is driven by the behavior of mergers.   Halos in the early stages of a merger will have a higher $\sigmadm$ at fixed mass than the mean relation.  If, during these merger events, the collisionless dark matter accretes faster than the baryons (since the extended \PH gas envelopes of accreting satellites will be more easily ablated during the merger encounter), then the ICM mass fraction will be locally depressed, driving an anti-correlation between $\sigmadm$ and $\ficm$.  We note that these effects are absent in the \GO case, since the correlation of that model is positive, $C = 0.28$.

The concentration and ICM emission measure are measures that are sensitive to formation history \citep{wechsler:02,busha:07}  and substructure driven by merging.  The correlations between these parameters and $\sigmadm$ are surprisingly modest, $\sim 0.2$, perhaps indicating that formation history is a more important driver, compared to recent mergers, for these measures.  There is not a strong sensitivity to gas physics in the $c-\sigmadm$  correlation, but the \PH $\hat{Q}-\sigmadm$ correlation of $0.32$ is significantly larger than the \GO value.  In fact, the only two signals with lower $\sigmadm$ correlations in \GO than in \PH are $\tsl$ and $\hat{Q}$, both measures that are sensitive to small regions of cool, dense gas.  

Both $Y$ and $\elbol$ have a weaker correlation with $\sigmadm$ in \PH than in \GO.  
As $Y \propto \ficm T_m$, we expect that we can describe the scatter in $Y$ at fixed mass as 
\begin{equation}
\sigma_Y^2 = \sigma_T^2 + \sigma_f^2 + 2C_{Tf} \sigma_T \sigma_f .
\label{eqn:sigmay}
\end{equation}
Examining Table~\ref{tab:quint}, we see that, because the scatter in $Y$ is largely dominated by $T_m$ in the \GO model and $\ficm$ in the \PH case, the value of the correlation coefficient, $C_{Tf}$, is not an important factor.  Our value for the scatter in $Y$ is higher than that measured in the simulations of \cite{kravtsov:06}, as they see 
an anti-correlation in the X-ray inferred deviations of gas mass and temperature at fixed mass.  In the intrinsic measures of Table~\ref{tab:r0}, we find positive correlation between $\ficm$ and $T_m$ with values of $0.42$ (\PH) and $0.48$ (GO).  

It is worth noting the interesting case of $C = 0.08$ between $Y$ and $\sigmadm$ in the \PH simulation.  Algebraically, this is unsurprising due to the positive correlation between $\sigmadm$ and $T_m$ and the negative correlation with $\ficm$.   This lack of correlation suggests that SZ surveys should produce cluster samples that unbiased with respect to dynamical state.  Similarly, the low correlation between $Y$ and $\hat{Q}$, $C = 0.12$, shows that SZ surveys should not be strongly biased by gas clumpiness.

When metals are ignored, the bolometric luminosity scaling, $L \propto \ficm^2 \hat{Q} T^{1/2}$, implies that the scatter in 
luminosity at fixed mass follows 
\begin{equation}
\sigma_L^2 = 4 \sigma_f^2 + \sigma_Q + \frac{1}{4}\sigma_T^2 + 4\Psi_{fQ} + 2\Psi_{fT} + \Psi_{QT} .
\label{eqn:sigmal}
\end{equation}
For the \PH case, the scatter in ICM mass fraction is the primary contributor, responsible for $90\%$ of the variation in $\elbol$.  In contrast, $\ficm$ variations account for only $7\%$ of the scatter in the \GO case, where the majority contributors are variations in $\tsl$ and $\hat{Q}$.   Our results are consistent with the work of \cite{balogh:06}, who use an analytic model to show that variation in halo structure cannot completely account for the observed variance in $\elbol$ at fixed mass.

In both simulations, the correlation between $T_m$ and $\tsl$ at fixed mass is very high.  Although a given halo will not have the same $T_m$ and $\tsl$, the high correlation coefficients indicate that the two temperature measures similarly trace the thermal state of a halo.  The correlation is higher in the \PH simulation than in the \GO, and cool cores in the latter model also drive higher correlations between $\tsl$ and $\hat{Q}$ than between $T_m$ and $\hat{Q}$.   The correlations between $\ficm$ and temperature measures are nearly constant between the two simulations, with values in the range $0.4-5$ for both $T_m$ and $\tsl$.  

The highest measures of correlation are between $T_m$ and $Y$ in the \GO case ($0.97$), and between $\ficm$ and $Y$ in the \PH run ($0.88$).   The SZ-X-ray correlation of $Y$ and $\elbol$ is also large, $\sim 0.7$, in both treatments.  This robust behavior is promising for cross-calibrations of future, combined X-ray and SZ surveys \cite{younger:06, cunha:08}. 

Both simulations also have a significantly non-zero correlation coefficient between $\elbol$ and $\tsl$, 
$0.73$ in \PH and $0.67$ in \GO.  As shown by \cite{nord:08}, the $L-T$ scaling relation expected from X-ray flux-limited surveys is sensitive to the value of $C_{TL}$, and studies of current and future samples will be able to place limits on this correlation, using techniques similar to those employed by \cite{rozo:09} to constrain the correlation of mass and X-ray luminosity at fixed optical richness for the SDSS \maxbcg\  sample.

%%%%%%%%%%%%%%%%%%%%%%%%%%%%%%%%%%%%%%%%%%%%%%%%%%%%%%%%%
\subsection{Redshift Evolution of the Signal Covariance}\label{sec:revol}

\begin{figure}
\plotone{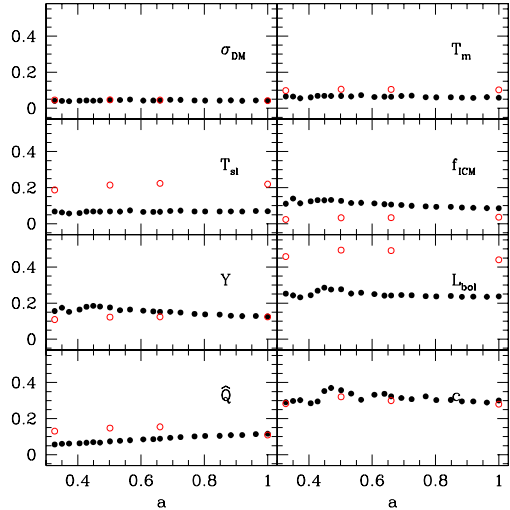}
\caption{Evolution of the scatter in eight bulk cluster properties.
\label{fig:scatevol}}
\end{figure}

We plot the time evolution of the signal scatter in Figure \ref{fig:scatevol}, with error bars come from bootstrap resampling.  
Going back to $z=2$, few halo properties show any evolution with redshift.  In the \PH case, the scatter in baryon fraction slightly increases near $z = 2$, causing an increase in the scatter of $Y$ and $L$.  On the other hand, the scatter in emission measure decreases at higher redshift.  Note that, at redshift two, there are only 62 halos in the \PH simulation 
above our mass cut of $5 \times 10^{13} \hinv \msun$.   At redshift zero, there is a higher scatter at the low-mass end of our mass range, as seen in Figure \ref{fig:scalestructure}.  Since the halos at redshift $z=2$ are only slightly over the mass cut, their higher scatter may reflect this mass dependence rather than pure redshift evolution.  

\begin{figure}
\plotone{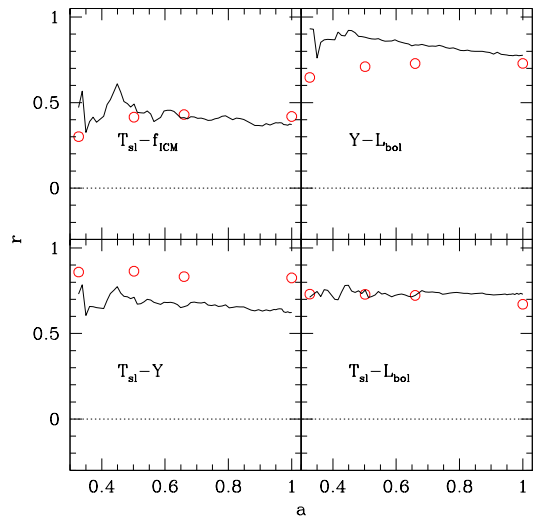}
\caption{Four pairs of signals which show little to no evolution in the correlation coefficient.  The 
black, solid line denotes the \PH simulation, and the red, open points the \GO simulation.
\label{fig:revol1}}
\end{figure}

Moving on to the off-diagonal elements, most pairs of signals show little evolution in the correlation coefficient 
with redshift.  Figure \ref{fig:revol1} shows four typical pairs with little evolution.  Even as 
the physical density of halos changes with redshift, we see that the interplay of $Y$ and $L$ or 
$\tsl$ and $\ficm$ does not change.   We see particularly little evolution in pairs of signals in the \PH 
simulation.

\begin{figure}
\plotone{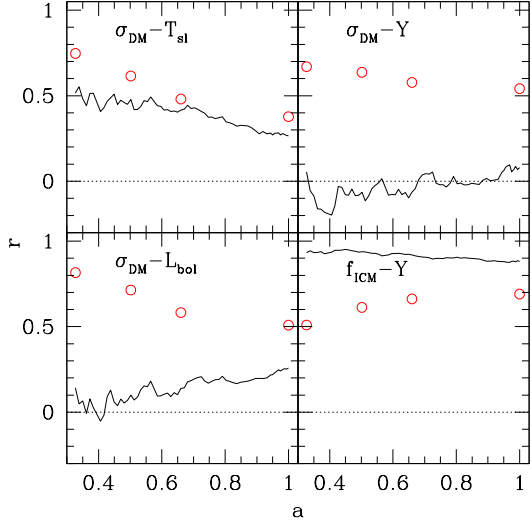}
\caption{Four pairs of signals which show some degree of evolution in the correlation coefficient, 
particularly in the \GO simulation.  The 
black, solid line denotes the \PH simulation, and the red, open points the \GO simulation.
\label{fig:revol2}}
\end{figure}

We do see evolution in a few signal pairs, notably between $\sigmadm$ and other signals in the \GO 
simulation.   Several pairs of signals -- such as $\sigmadm-\tsl$ and 
$\ficm-Y$ -- evolve in the \GO simulation, but not in the \PH simulation.  
Furthermore, for $\sigmadm-\elbol$, we see evolution in both simulations, but in opposite directions.   The 
evolution of these pairs is shown in Figure \ref{fig:revol2}.

The correlation between $\sigmadm$ and the SZ and X-ray signals increases towards high redshift in the 
\GO simulation.  We speculate that this is due to the increase in $\hat{Q}$ with time at fixed mass in the \GO simulation, as shown in 
Figure \ref{fig:scalexray}.  The gas in the \GO simulation develops more substructure with time, and this tends to decouple the SZ and X-ray measurements from the dark matter velocity dispersion, $\sigmadm$. 
Although $\hat{Q}$ also increases with time in the \PH simulation, its normalization at all redshifts is 
much lower than in the \GO simulation.  Hence, the substructure of the gas does not contribute much to the 
$\tsl$ or $L$ measures in the \PH simulation, and there is little evolution of their correlation 
with $\sigmadm$.  

%%%%%%%%%%%%%%%%%%%%%%%%%%%%%%%%%%%%%%%%%%%%%%%%%%%%%%%%%%%%%%%%%%%%%
\subsection{Implications for Multi-Signal Mass Selection}\label{sec:imp}

At fixed signal, the scatter in halo mass is $\sigma_{\mu i} = \sigma_i/ \alpha_i$, where $i$ labels the particular signal and $\alpha_i$ is the slope of that signal--mass relation.  From the analysis above, we compute the mass scatter for a subset of  $z=0$ signals and present the data in the first two columns of Table \ref{tab:masscat}.   We see that $Y$ provides the best mass selection under both physical treatments, with scatter $\sim 7\%$.  For the \PH case, the X-ray luminosity is quite good, with a $10\%$ scatter, while $\ficm$ is the worst selector, with the weak $\ficm-M$ slope producing a $28\%$ scatter in mass.  We will see below that this large scatter can actually be used to improve selection when paired with more precise, correlated signals.  

When selecting halos using multiple signals, the mass variance is $\Sigma^2  \ = \  ( \mathbf{\alpha}^\dagger \Psi^{-1} \mathbf{\alpha} )^{-1}$, where $\alpha$ is a vector of slopes with elements $\alpha_i$.  
We consider the two-signal case, and let $r \equiv C_{12}$ be the (traditional) correlation coefficient between signals 1 and 2.  
% The signal covariance can be written as 
% \begin{equation}
% \Psi = \left(
% \begin{array}{cc}
% \sigma_1^2 & r\sigma_1\sigma_2 \\
% r\sigma_1\sigma_2 & \sigma_2^2
% \end{array}
% \right).
% \end{equation}

\begin{figure}
\plotone{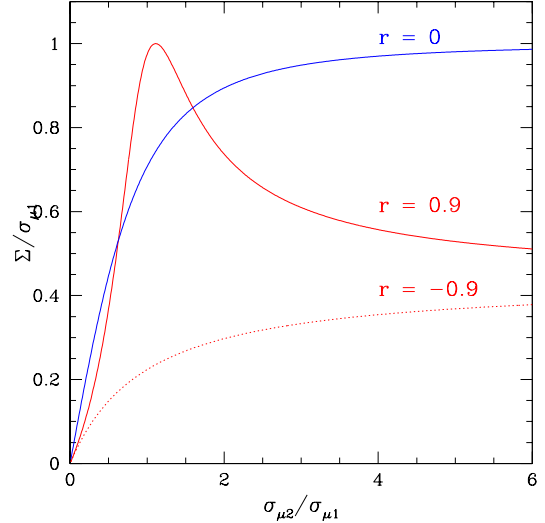}
\caption{For a range of correlation coefficients, we plot the dispersion in mass achieved by combining a pair of signals (1 and 2) against the ratio of their dispersions.  Different lines vary the correlation coefficient $r$ between the pair of signals at fixed mass.  Given a measurement of the first signal, a second measurement always improves the mass scatter, except in the degenerate case, $\sigma_{\mu 2}/\sigma_{\mu 1} =1$ and $r=1$.   While negatively correlated signals yield more improved mass selection than positively correlated pairs, the improvement becomes independent of sign when the second signal is much noisier than the first, $\sigma_{\mu 2}/\sigma_{\mu 1} \gg 1$.  
\label{fig:massr}}
\medskip
\end{figure}

Given a pair of signal measurements, the mass selection is log-normal with variance 
\begin{equation}\label{eqn:masscat}
\Sigma^2 = (1-r^2) ( \sigma_{\mu 1}^{-2} + \sigma_{\mu 2}^{-2} -2r\sigma_{\mu 1}^{-1} \sigma_{\mu 2}^{-1})^{-1}.
\end{equation}
The improvement in mass selection to be gained by a pair of signals, relative to a single measurement, is displayed in Figure \ref{fig:massr}.  Here, we plot $\Sigma/\sigma_{\mu 1}$ as a function of the scatter ratio ($\sigma_{\mu 2}/\sigma_{\mu 1}$) for several values of the correlation coefficient.    If the second signal is a better mass proxy, $\sigma_{\mu 2} \ll \sigma_{\mu 1}$,  then it dominates the selection. Combining signals with comparable mass selection, $\sigma_{\mu 2} \sim \sigma_{\mu 1}$,  can result in anything from no improvement in the degenerate case ($r \rightarrow 1$) to the $\sqrt{2}$ improvement for uncorrelated signals ($r=0$) to dramatic improvement in the anti-correlated case ($r \rightarrow -1$).   Surprisingly, when an intrinsically ``noisy'' mass proxy is added, $\sigma_{\mu 2} \gg \sigma_{\mu 1}$, one still achieves significant improvement in mass selection as long as the signal pair correlation is large in an absolute sense.

\begin{deluxetable*}{l|ccccc}
\tablecaption{Mass Scatter at Redshift Zero \label{tab:masscat}\tablenotemark{a}}
\tablehead{
\colhead{Signal}  & \colhead{$\sigmadm$} & \colhead{$\tsl$} & \colhead{$\ficm$} & \colhead{$Y$} & \colhead{$\elbol$}   }
\startdata
$\sigmadm$   & $0.12 \backslash 0.12$  & $0.12$ & $0.12$ & $0.075$ & $0.12$ \\
$\tsl$   & $0.10$ &  $012 \backslash 0.38$  & $0.35$ & $0.050$ & $0.26$ \\
$\ficm$   & $0.11$ & $0.12$ &$ 0.28 \backslash 0.12$  & $0.054$ & $0.21$ \\
$Y$ & $0.062$ & $0.069$ & $0.041$ &  $0.069 \backslash 0.075$  & $ 0.066 $ \\
$\elbol$ & $0.09$ & $0.10$ & $0.09$ & $0.069$ & $0.10 \backslash 0.26$   
\enddata
\tablenotetext{a}{The diagonal elements give the mass scatter for individual signals for the $\PH\backslash\GO$ cases.  Off-diagonal elements give the halo mass scatter for signal pairs, with the \PH case in the lower triangle and the \GO case in the upper, as in Figure \ref{fig:covarz0} .}
\end{deluxetable*}

We evaluate equation~(\ref{eqn:masscat}) for a set of observable signal pairs and present the resultant mass scatter values in the right-hand columns of Table \ref{tab:masscat}.  In the \PH simulation, the correlation coefficient between $L$ and $Y$ is $r = 0.78$, and the scatter in mass for the pair of signals is $6.9\%$, a mild improvement over the scatter using only $\elbol$ ($11\%$) or only $Y$ ($7.5\%$).  Given the high covariance between $\elbol$ and $Y$, future multiwaveband surveys that join SZ and X-ray detections should have a strong mass selection properties.

Finally, we note that the best mass selection comes in the \PH case from combining $Y$ with $\ficm$.  The combination of a strong correlation, $r=0.88$, and, especially, the relatively large degree of scatter in mass at fixed $\ficm$ ($0.28$) produce a scatter in mass of only $4.1\%$ when measurements of these two signals are joined.  We caution that this analysis does not take measurement errors into account, and it is clear that high data quality will be important to realize this level of mass selection.  

%%%%%%%%%%%%%%%%%%%%%%%%%%%%%%%%%%%%%%%%%%%%%%%%%%%%%%%%%%%%%%%%%%%%%
%%%%%%%%%%%%%%%%%%%%%%%%%%%%%%%%%%%%%%%%%%%%%%%%%%%%%%%%%%%%%%%%%%%%%
\section{Conclusion}\label{sec:con}

We analyze scaling relations for multiple properties of massive halos taken from a pair of gas dynamic simulations with  
different gas physics treatments.   Our samples contain tens of thousands of halos with masses $M_{200} > 5 \times 10^{13} \hinv \msol$ at redshifts $z \lta 2$.  The physical treatments of gravity only ($\GO$) or preheating ($\PH$) are both highly idealized, but we show that the latter reproduces the scaling relation behavior of core-extracted X-ray measures of local clusters.   

The dark matter velocity dispersion scales with mass and redshift according to self-similar expectations, indicating that the virial theorem is respected regardless of gas treatment.  However, the gas behavior in both treatments differs from self-similarity.   The deviations in the \GO case tend to be small and are related to the mass-limited sample definition.  
% Mass-weighted gas temperatures at high redshift are lower than expectations based on self-similarly extrapolating the $z=0$ scaling, but this effect is consistent with enhanced merger activity driving larger turbulent gas motions in the high-$z$ sample.  When this kinetic energy is added to the gas thermal energy, the combined total energy obeys self-similar expectations at the percent level.  

The entropy injection of the \PH model drives more substantial deviations from self-similar scaling.  At $z=0$, the ICM mass fraction varies with mass in a manner roughly consistent with observed measurements of local clusters.   We find that $\ficm$ requires a quadratic fit in log-mass, and mild curvature in the logarithmic scalings of $Y$ and $\elbol$ as a function of mass is also evident.  While the ICM in \PH halos is lower in mass compared to the \GO case, it is also slightly hotter.   The effects on $\ficm$ and $T_m$ nearly cancel when combined to form the thermal SZ signal, leading to nearly identical $Y$--$M$ scaling relations above $3 \times 10^{14}$ for both \PH and \GO cases at low redshift.   

The ICM mass fraction at fixed mass declines weakly with redshift, by $10\%$ in $5 \times 10^{14} \hinv\msol$ halos at $z=1$ and with larger reductions at lower masses.  While {\sl Chandra\/} observations of optically-selected clusters in the RCS survey show evidence of reduced gas fractions at $z=1$ \citep{hicks:06}, further work is needed to address the quantitive level of agreement between the observations and \PH model expectations.     
The \PH baryon fraction evolution drives departures from self-similar predictions in the  $Y$--$M$ and $\elbol$--$M$ relations;  slopes tend to steepen slightly and the normalization at $10^{14} \hinv\msol$ is lower than self-similar expectations at high redshift.   

We present the first systematic investigation of property covariance in the computational samples of massive halos.  The data generally support a multivariate log-normal form for the joint distribution of signals at fixed mass and redshift.   All measures depart somewhat from  an exact gaussian form, but the deviations in gas measures are smaller for the \PH model due to the suppression of substructure caused by the preheating.   Most signal pairs exhibit positive correlations, with the lone exception of $-0.1$ between $\sigmadm$  and $\ficm$ in the \PH case.  The thermal SZ signal displays a robust $13\%$ scatter that is strongly correlated with variations in both ICM gas mass and temperature, with $\ficm$ dominating in the \PH case and $T_m$ being more important in the \GO treatment.   

Combining multiwavelength observations offers an opportunity to improve selection of clusters by their intrinsic mass.   We derive the mass variance of signal pairs and show that combining strongly correlated signals always improves mass selection, even when one of the signals by itself is a comparatively poor mass proxy.   The combination of thermal SZ and ICM mass fraction in the \PH case selects halo mass with just $4\%$ intrinsic rms scatter.   

Identifying the root causes behind the terms in the covariance matrix is a considerable task that we leave for future work.   Mergers \citep{roettigerBurnsLoken:97, rickerSarazin:01}  and assembly bias (\cite{Boylan-Kolchin:09} and references therein) will surely play important roles for many cluster signals.  Studies of merger history behavior will shed light on survey selection properties, particularly potential biases related to the dynamical state of a halo.   

The simple treatment of baryon physics in our simulations limits our investigation to the hot, thermal ICM of clusters.  More extensive physical treatments that incorporate galaxy and supermassive black hole formation and other physics such as MHD and non-thermal plasmas will ultimately extend the set of observable halo signals into the optical/NIR and radio wavebands.

%%%%%%%%%%%%%%%%%%%%%%%%%%%%%%%%%%%%%%%%%%%%%%%%%%%%%%%%%%%%%%%%%%%%%
\acknowledgements
AEE acknowledges support from NSF AST-0708150.  Support for this work was provided by NASA through Chandra Postdoctoral Fellowship grant number PF6-70042 awarded by the Chandra X-ray Center, which is operated by the Smithsonian Astrophysical Observatory for NASA under the contract NAS8-03060.
%thank you....
%%%%%%%%%%%%%%%%%%%%%%%%%%%%%%%%%%%%%%%%%%%%%%%%%%%%%%%%%%%%%%%%%%%%%
% \appendix
% \section{Appendix}\label{sec:appendix}

%%%%%%%%%%%%%%%%%%%%%%%%%%%%%%%%%%%%%%%%%%%%%%%%%%%%%%%%%%%%%%%%%%%%%
\newpage
\bibliographystyle{apj}
%\bibliography{bibfile}

%\bibliography{mastercovar}

%%%%%%%%%%%%%%%%%%%%%%%%%%%%%%%%%%%%%%%%%%%%%%%%%%%%%%%%%%%%%%%%%%%%%
%%%%%%%%%%%%%%%%%%%%%%%%%%%%%%%%%%%%%%%%%%%%%%%%%%%%%%%%%%%%%%%%%%%%%

%%%%%%%%%%%%%%%%%%%%%%%%%%%%%%%%%%%%%%%%%%%%%%%%%%%%%%%%%%%%%%%%%%%%%%
%%%%%%%%%%%%%%%%%%%%%%%%%%%%%%%%%%%%%%%%%%%%%%%%%%%%%%%%%%%%%%%%%%%%%%
%Figures go below here
%%%%%%%%%%%%%%%%%%%%%%%%%%%%%%%%%%%%%%%%%%%%%%%%%%%%%%%%%%%%%%%%%%%%%%

\end{document}